\providecommand{\columnwiselinenumbersfalse}{}%
\providecommand{\runningpagewiselinenumbers}{}%
\PassOptionsToPackage{desactivate}{linenoaa}
\documentclass{aa}
\usepackage{graphicx}
\usepackage{txfonts}
\usepackage{lipsum}
\usepackage{lscape}
\usepackage{placeins}
\usepackage{amsmath, amssymb, mathrsfs}
\usepackage{booktabs}
\usepackage{listings}
\usepackage{anyfontsize}
\usepackage[svgnames]{xcolor}
\usepackage{hyperref}
\usepackage{orcidlink}
\let\CheckCommand\providecommand
\usepackage{microtype}

\definecolor{C0blue}{RGB}{31,119,180}
\definecolor{C1orange}{RGB}{255,127,14}
\definecolor{C2green}{RGB}{44,160,44}
\definecolor{C3red}{RGB}{214,39,40}
\definecolor{C4purple}{RGB}{148,103,189}
\definecolor{C5brown}{RGB}{140,86,75}
\definecolor{C6pink}{RGB}{227,119,194}
\definecolor{C7gray}{RGB}{127,127,127}
\definecolor{C8olive}{RGB}{188,189,34}
\definecolor{C9cyan}{RGB}{23,190,207}

\hypersetup{
    breaklinks=true,
    colorlinks=true,
    linkcolor=C2green,
    citecolor=C0blue,
    filecolor=magenta,      
    urlcolor=C9cyan,
    pdftitle={CHEPS EMPEROR - Pe{\~n}a, P. et al},
    pdfauthor={Pe{\~n}a, Pablo},
    }

\newcommand{\vsini}{$v\sin{i}$}

\newcommand{\ms}{ms$^{-1}$}
\newcommand{\kms}{kms$^{-1}$}

\newcommand{\msini}{m~sin~$i$}
\newcommand{\emp}{\texttt{EMPEROR}}

\newcommand{\reddemc}{\texttt{reddemcee}}

\newcommand{\ins}{\text{INS}}

\newcommand{\Uniform}[2]{$\sim$$\mathcal{U}(#1, #2)$}
\newcommand{\Normal}[2]{$\sim$$\mathcal{N}(#1, #2)$}
\newcommand{\Isotro}[2]{$\sim$$\mathcal{I}(#1, #2)$}

\newcommand{\lnzh}{$\ln{\widehat{\mathcal{Z}}}$}

\newcommand{\zerok}{$0\mathcal{K}$}
\newcommand{\onek}{$1\mathcal{K}$}
\newcommand{\twok}{$2\mathcal{K}$}

\newcommand{\logrhk}{$\log{R'_{\mathrm{HK}}}$}
\newcommand{\smw}{$S_{\mathrm{MW}}$}

\newcommand{\refsec}[1]{\hyperref[#1]{\S\ref*{#1}}}
\newcommand{\refeq}[1]{\hyperref[#1]{Eq.\ref*{#1}}}
\newcommand{\reffig}[1]{\hyperref[#1]{Fig.\ref*{#1}}}
\newcommand{\reftab}[1]{\hyperref[#1]{Table \ref*{#1}}}
\newcommand{\refapp}[1]{\hyperref[#1]{Appendix \ref*{#1}}}

\begin{document}
    \title{Cold giant discoveries from a joint radial-velocity and astrometry framework}

    \subtitle{EMPEROR II -- Jupiter analogues around metal-rich CHEPS stars}

    \author{Pablo ~A. Pe{\~n}a R.\orcidlink{0000-0002-8770-4398}\inst{1,2} \and
    James S. Jenkins\orcidlink{0000-0003-2733-8725}\inst{1,2} \and
    Fabo Feng\orcidlink{0000-0001-6039-0555}\inst{3} \and
    Douglas R. Alves\orcidlink{0000-0002-5619-2502}\inst{1,2} \and
    Florence de Almeida\orcidlink{0009-0002-5653-8814}\inst{1} \and
    Fr\'ed\'eric Dux\orcidlink{0000-0003-3358-4834}\inst{4} \and
    Guang-Yao Xiao\orcidlink{0000-0001-6753-4611}\inst{3} \and
    Joanne M. Rojas M.\orcidlink{0009-0007-6262-1194}\inst{1,2} \and
    Jose I. Vines\inst{5} \and
    Rafael I. Rubenstein\orcidlink{0000-0002-2460-0305}\inst{1,6} \and
    R. Ram\'irez Reyes\orcidlink{0000-0002-4678-8361}\inst{2, 7} \and
    Suman Saha\orcidlink{0000-0001-8018-0264}\inst{1,2} \and
    Connor J. Cheverall\orcidlink{0009-0005-4330-7197}\inst{1,2} \and
    Mat\'ias R. D\'iaz\orcidlink{0000-0002-2100-3257}\inst{8}
        }

    \institute{Instituto de Estudios Astrof\'isicos, Facultad de Ingenier\'ia y Ciencias, Universidad Diego Portales, Av. Ej\'ercito 441, Santiago, Chile
            \and Centro de Astrof\'isica y Tecnolog\'ias Afines (CATA), Casilla 36-D, Santiago, Chile
            \and State Key Laboratory of Dark Matter Physics, Tsung-Dao Lee Institute \& School of Physics and Astronomy, Shanghai Jiao Tong University, Shanghai 201210, People’s Republic of China
            \and
            Institute of Physics, Laboratory of Astrophysics, \'Ecole Polytechnique F\'ed\'erale de Lausanne (EPFL), Observatoire de Sauverny, 1290 Versoix, Switzerland 
            \and Instituto de Astronom\'ia, Universidad Cat\'olica del Norte, Angamos 0610, 1270709, Antofagasta, Chile
            \and United States Fulbright Fellow; Chile Fulbright Commission
            \and Instituto de Astrof\'isica, Departamento de F\'isica y Astronom\'ia, Facultad Ciencias Exactas, Universidad Andr\'es Bello, Fern\'andez Concha 700, Las Condes, Santiago, Chile
            \and Las Campanas Observatory, Carnegie Institution for Science, Raul Bitr\'an 1200, La Serena, Chile
            }

    \date{}

    \abstract{The population of long-period giant planets shapes planetary system architectures and formation pathways, but these `cold-Jupiters' remain relatively unexplored. Radial velocity (RV) surveys lose sensitivity at multi-AU separations and provide only minimum planet masses, while transit surveys have poor detection probability at long periods. Absolute astrometry from the Hipparcos and Gaia missions offer an additional source for stellar motion that can break the orbital inclination degeneracy and strengthen detection confidence. This is especially timely ahead Gaia DR4/DR5, expected to enable routine astrometric vetting and true-mass measurements for long-period RV planets.} 
    {Extending the Chile-Hertfordshire ExoPlanet Survey (CHEPS) by combining RVs spanning up to 16 years with absolute astrometry, we search for and characterise cold giants around metal-rich FGK stars.} 
    {We upgrade the \emp~framework, incorporating astrometric differencing to jointly fit RVs and astrometry for five CHEPS targets, performing Bayesian model comparison and baseline- and phase-coverage metrics to quantify the astrometric contribution.} 
    {Our analysis confirms and characterises orbital parameters for two known planets in HIP\,21850 and detects five new planets: a warm Jupiter--HIP\,10090c, with orbital period $P=321.8^{+0.3}_{-0.6}$~d and mass $M=0.85^{+0.03}_{-0.12}~M_J$, and four Jupiter analogues--HIP\,8923b with $P=14.1^{+0.4}_{-0.7}$~yr and $M=9.98^{+0.78}_{-0.16}~M_J$, HIP\,10090b with $P=8.1^{+0.3}_{-0.3}$~yr and $M=3.87^{+0.65}_{-0.60}~M_J$, HIP\,39330b with $P=12.7^{+0.6}_{-0.7}$~yr and $M=1.68^{+0.16}_{-0.13}~M_J$, and HIP\,98599b with $P=7.3^{+0.1}_{-0.1}$~yr and $M=6.85^{+0.10}_{-0.22}~M_J$. Adding astrometry reduces period and mass uncertainties by factors between 3 and 10, whilst increasing the Bayes' factors by up to $\sim$60.}
    {The synergy of long-baseline RVs and absolute astrometry provides a robust pathway to discover and characterise cold giant planets and metal-rich Solar-System analogues. Our results demonstrate that astrometric samples meaningfully improve detection confidence and convert minimum masses into true masses. This approach fills the demographic gap between RV and direct imaging sensitivities and prepares the field for forthcoming astrometric missions that will allow to find stellar systems like our own.} 


   \maketitle

\section{Introduction}  \label{sec:intro}

Metal-rich stars have played a central role in exoplanet discovery since the earliest gas giant detections, and the planet-metallicity correlation remains one of the most robust demographic trends: gas giants are markedly more common around metal-rich FGK stars \citep{1997MNRAS.285..403G, 2004A&A...415.1153S, 2005ApJ...622.1102F, 2018haex.bookE.195W}. Today, there are over six thousand confirmed exoplanets, the vast majority found by radial velocities (RV) and transit photometry. However, the sensitivity of both techniques declines with orbital separation $a$ \citep{2021ARA&A..59..291Z}: the geometric transit probability scales as $R_\star/a$, and long-period RV signals require many years of high-precision monitoring to resolve low-amplitude, slowly varying signals \citep{2008PASP..120..531C}. As a result, the population of cold-orbiting giants beyond a few AU remains comparatively unexplored.

In the rest of this manuscript, when referring to exoplanets we adopt the term `cold' for those with semi-major axis of 1--10~AU, `wide-orbiting' for >10~AU, and `Jupiter analogue' for 3--7~AU and mass 0.3--13~$M_J$ \citep{2019AJ....157...52B}. Recent occurrence-rate analyses show that Jupiters with orbital periods >100~d occur around Sun-like stars at 6--7\%, nearly an order of magnitude more common than hot Jupiters \citep{2020MNRAS.492..377W}. Moreover, the relative occurrence of hot to cold Jupiters appears to decrease with stellar mass, consistent with expectations from classical core accretion and migration timescales \citep{2024ApJ...967...74G}. These trends highlight the need for long-baseline surveys capable of detecting and characterising cold giants.

While precise RVs constrain orbital parameters and minimum masses $M_{\mathrm{pl}}\sin{i}$, they do not measure inclination and thus cannot yield true masses by themselves. Absolute astrometry, on the other hand, is directly sensitive to the sky-projected stellar reflex motion, and provides dynamical masses and inclinations, albeit with sensitivity that degrades with distance $d$. In particular, the signal-to-noise ratio (SNR) scales as $\mathrm{SNR_{AM}} \propto \frac{M_{\mathrm{pl}}\cdot a }{M_\star \cdot d}$, while for RVs this ratio scales as $\mathrm{SNR_{RV}} \propto \frac{M_{\mathrm{pl}}}{\sqrt{a \cdot M_\star}}$.

Combining RVs with absolute astrometry breaks key degeneracies, enhances SNR, and enables robust characterisation of long-period companions even when a single technique samples only a fraction of the orbit. With Hipparcos and Gaia providing high-precision positions, proper motions, and accelerations over a $\sim$24~yr baseline, the RV+astrometry synergy has matured to a powerful pathway to measure true masses of cold giants and identify precise dynamical mass for Jupiter analogues \citep{2014ApJ...797...14P, 2022A&A...657A...7K, 2019MNRAS.490.5002F, 2021ApJS..254...42B}.

A paradigm shift is imminent within the precision-RV community: long-period RV candidates will increasingly be treated as testable astrometric hypotheses and systematically checked for consistency with Gaia. In practice, an RV posterior predicts an on-sky reflex signal (amplitude and orientation) that Gaia can confirm, refute, or tighten into a true-mass measurement. Recent forecasts suggest that Gaia astrometry may yield 7\,500 ± 2\,100 planet detections in DR4 and 120\,000 ± 22\,000 in DR5, with $\sim$1\,900 ± 540 (DR4) and 38\,000 ± 7\,300 (DR5) having masses and periods measured to be better than 20\%. Most of these detections are expected to be super-Jupiters on $\sim$2--5~AU orbits around nearby GKM stars \citep{2026AJ....171...18L}, overlapping with the parameter space targeted by long-baseline Doppler surveys.

The connection between outer cold giants and inner small planets is crucial to understanding system architectures. Early observational work reported that systems with inner super-Earths frequently host cold giants (roughly one every three) implying a positive correlation \citep{2018ApJ...860..101Z, 2019AJ....157...52B}. Subsequent studies found lower conditional rates or no excess in carefully vetted samples \citep{2023A&A...677A..33B}. A more nuanced, mass-dependent picture is emerging: new analyses indicate that Saturn-mass cold giants are positively correlated with inner super-Earths, whilst the most massive super-Jupiters are not, potentially reconciling previous discrepancies and informing models of accretion efficiency and dynamical histories in multi-planet systems \citep{2025ApJ...988..101L}. Establishing how common true Jupiter analogues are, and measuring their dynamical masses, bears directly on how typical the Solar System is among planetary systems.

Cold giants regulate both assembly and survival of inner planetary systems, sculpt debris belts, and provide boundary conditions for planet formation theories \citep[migration histories, core accretion][]{2013ApJ...775...42I, 2012A&A...541A..97M}. Measuring their occurrence, mass function, and orbital architectures tests these theories where they differ most, informs the demographics accessible to direct imaging, and provides prime-targets for follow-up. Looking ahead, the RV-astrometry coupling is explicitly built into future facilities: the Tianyu transit project plans confirmation and characterisation of long-period giants via Gaia astrometry and RV follow-up, leveraging multi-survey baselines \citep{2024AcASn..65...34F}; and Habitable Worlds Observatory \citep[HWO, ][]{2024AAS...24421004D} studies emphasize that precursor RVs together with a modest number of astrometric epochs can deliver full orbital solutions for cold giants and habitable-zone planets, enabling efficient direct-imaging campaigns \citep{2025AJ....170..208S}.

The Chile-Hertfordshire ExoPlanet Survey (CHEPS) was initiated in 2009 to target inactive, metal-rich stars with high-precision RVs from both HARPS and CORALIE, explicitly leveraging the planet-metallicity correlation to find gas giants efficiently \citep{2008A&A...485..571J, 2009MNRAS.398..911J, 2017MNRAS.466..443J}. By extending the CHEPS observations through 2025, we assemble a 16-year time baseline (2009--2025) ideally suited to unveil long-period curvatures and complete (or near complete) orbits for cold exoplanets. In this paper, we leverage this long CHEPS baseline with modern astrometric constraints. We (1) identify five CHEPS targets exhibiting long-term RV curvature (HIP 8923, HIP\,10090, HIP\,21850, HIP\,39330, and HIP\,98599); (2) model each system with a joint RV+astrometry framework using \emp~\citep{emperor-paper}; and (3) determine the true masses and orbital parameters of the suspected cold giants. Beyond target-level results, our analysis serves as an end-to-end validation for the \emp~approach for combined RV+astrometry inference on long period planets. The manuscript is organised as follows: \refsec{sec:emperor} outlines the joint modelling framework; \refsec{sec:observations} summarises datasets and stellar parameters; \refsec{sec:results} presents system-by-system fits; \refsec{sec:discussion} discusses and quantifies the benefits of the joint modelling, contextualises the discovered systems into the overall exoplanetary census, and addresses possible drawbacks in this study; and \refsec{sec:conclusions} summarises the key points of this research.

\section{Overview of the modelling framework}  \label{sec:emperor}

\emp~\citep[Exoplanet Mcmc Parallel tEmpering for Rv Orbit Retrieval, ][]{emperor-paper} is a modular, Python-based pipeline that searches RV time series for planetary signals and performs Bayesian model comparison to select the preferred solution. Posteriors are sampled with \reddemc~\citep{reddemcee-paper}, an adaptive parallel tempering MCMC sampler \citep[APT, ][]{2005PCCP....7.3910E, 2016MNRAS.455.1919V} designed to explore broad multi-modal posteriors efficiently and mitigate local trapping, significantly accelerating convergence rates while improving the reliability of global inferences. The framework estimates the marginalised likelihood--also known as evidence--through a hybrid approach that combines thermodynamic integration \citep{Gelman1998SimulatingNC} and stepping stones \citep{steppingstones2011} with realistic uncertainties. \emp~supports multi-instrument, multi-planet analyses, and flexible noise models including (when needed) stellar activity correlations and Gaussian processes. In what follows we summarise the Bayesian formulation and models used, and describe how we incorporate the Gaia-Hipparcos differencing framework into the joint likelihood \citep{2019MNRAS.490.5002F}.

\subsection{Bayesian formulation}\label{sec:bayesian_framework_bayesian_approach}
Bayes' theorem defines the posterior $p(\pmb{\theta} \mid D, M)$ of parameters $\pmb{\theta}$, given data $D$ and model $M$ as

\begin{equation} \label{eq:bayes_theorem}
    p(\pmb{\theta} \mid D, M) = \frac{p(\pmb{\theta} \mid M) \cdot p(D \mid \pmb{\theta}, M)}{p(D \mid M)}\,,
\end{equation}

where $p(\pmb{\theta} \mid M)$ is the prior, $p(D \mid \pmb{\theta}, M)$ the likelihood, and $p(D | M)$ the Bayesian evidence normalising the posterior \citep{robert_2007, 2019arXiv190508737F, 2022arXiv220211678L}. The evidence enables objective model comparison via Bayes' factors. In our case the likelihood is joint, coupling RVs and astrometry (subscript AM) through the shared orbital parameters $\pmb{\theta}$:
\begin{equation}
    p(D \mid \pmb{\theta},M) = p(D_{\mathrm{RV}} \mid \pmb{\theta}, M) \times p(D_{\mathrm{AM}} \mid \pmb{\theta}, M)\,.
\end{equation}

\subsection{Radial velocity model}\label{sec:frameworl-rvs}
The deterministic RV signal is a sum of Keplerian curves:

\begin{equation} \label{eq:keplerian}
    \mathcal{K}(t) = \sum_{j=1}^{N_{\mathrm{pl}}} K_j \cdot [\cos(\nu_j(t) + \omega_j) + e_j\cos(\omega_j)]\,,
\end{equation}

where each planet $j$ has a period $P_j$, semi-amplitude $K_j$, eccentricity $e_j$, longitude of periastron $\omega_j$, time of periastron passage $T_{0,j}$ and true anomaly $\nu_j$. Instrumental zero-points $\gamma_{0, \ins}$, long-term trends $\dot\gamma$, and (optionally) linear correlations with stellar activity indices $\mathcal{C}_{\mathcal{A},\ins} \cdot \mathcal{A}_{i,\ins}$--with $\mathcal{A}_{i,\ins}$ denoting the measured activity index and $\mathcal{C}_{\mathcal{A},\ins}$ the coefficient--are included additively into \refeq{eq:keplerian}. Two orbital angles remain unconstrained by RVs alone, the inclination $i_j$ (tilt with respect to the plane-of-sky) and the longitude of ascending node $\Omega_j$ (plane-of-the-sky orientation). Consequently, RVs measure only the minimum mass $M_{\mathrm{pl}, j}\sin i_j$ via the mass function

\begin{equation} \label{eq:mass-function}
    \frac{P_jK_j^3\sqrt{(1-e_j^2)^3}}{2\pi G} = \frac{(M_{\mathrm{pl}, j}\cdot\sin{i_j})^3}{(M_{\mathrm{pl}, j}+M_\star)^2}\,.
\end{equation}

\subsection{Astrometric Model (Hipparcos-Gaia differencing with IAD+GOST)}\label{sec:framework-astrometry}

We adopt the astrometric differencing approach of \citet{2019MNRAS.490.5002F, 2023MNRAS.525..607F, 2024ApJS..271...50F}. Rather than relying only on catalogue-level proper motion differences, we (i) fit the Hipparcos intermediate astrometric data \citep[IAD, the along-scan measurements at $\sim$1990--1993,][]{2007A&A...474..653V}; and (ii) propagate and project our model at the Gaia reference epoch using scan angles from the Gaia Observation Forecast Tool \citep[GOST, ][]{GOST_Web, GOST_Manual_2016}. This IAD+GOST approach preserves the Hipparcos epoch information and properly matches Gaia's along scan sensitivity, avoiding the information loss that can occur when collapsing to catalogue-level proper motion anomalies.

We propagate a five-parameter astrometric model between Hipparcos and Gaia epochs $(\Delta \alpha,\,\Delta \delta,\,\Delta \mu_{\alpha},\,\Delta \mu_{\delta},\,\Delta \varpi)$, including perspective terms from catalogue RV and parallax. We add the star's 3D reflex motion due to the companion(s) and project it onto the sky. Internally we use Thiele-Innes constants to map Keplerian elements to sky-plane displacements \citep[for a detailed description, see ][]{2024MNRAS.534.2858X}. For every Hipparcos scan we compute the predicted along-scan displacement from the model (barycentre + reflex) at the scan time and angle, and evaluate a Gaussian likelihood using the IAD uncertainties plus an additive jitter term $J_\mathrm{hipp}$. On the other hand, at the Gaia reference epoch we project the model into Gaia's along-scan direction using the GOST angles and parallax factors, and form the position and proper-motion differences relative to the Hipparcos solution linearly propagated to the Gaia epoch. The astrometric likelihood corresponds to the product of the Hipparcos-IAD term and the Gaia difference term. We include per-catalogue offsets and small frame-rotation jitters.

After removing linear-space motion, a companion induces a sky-plane acceleration $g$ that leaves two signatures over the Hipparcos-Gaia epoch separation $\Delta t$,
\begin{equation}
    \Delta r \approx \tfrac{1}{2} g(\Delta t)^2\,,\qquad \Delta\mu \approx g\,\Delta t\,,
\end{equation}
so jointly modelling positional ($\Delta r$) and proper-motion ($\Delta \mu$) offsets is especially sensitive to long-period companions. Using the actual scan geometries (GOST) reduces susceptibility to scan-angle–dependent systematics \citep[e.g.,][]{2023A&A...674A..25H}. Then, in this framework the GOST model corresponds to the geometric projection layer that converts the orbital model in sky coordinates into Gaia's effective along-scan observable, using GOST scan geometry, while the GDR2/3 model corresponds to the statistical likelihood layer that compares that predicted mean vector to the catalogue-level Gaia solution using the reported covariance matrix. Utilising multi-epoch Gaia astrometry (GDR2 and 3) can help differentiating between prograde and retrograde orbits, breaking the inclination degeneracy \citep{2024MNRAS.534.2858X}.

\subsection{Priors} \label{sec:framework-priors}

Our prior parameters are purposely flat for most RV and astrometric parameters. For long-period giants the empirical eccentricity distribution is broad \citep{2013MNRAS.434L..51K, 2025MNRAS.539..727S}, but to regularise poorly constrained cases and to avoid bias towards spuriously high $e$ in low SNR regimes, we use a truncated normal prior \Normal{0}{0.3} $\in [0, 1]$, penalising extremely higher eccentricities but allowing them if supported by the data \citep[see][]{2013A&A...549A..48T}. RV jitters use a wide truncated normal centred at zero, with the scale set by the average precision of the instrument, at either $\sigma=5$ or 10~\ms. Hipparcos additive astrometric jitter $J_\mathrm{hipp}$\Normal{1}{3} and fractional Gaia jitter $S_{\mathrm{gaia}}$\Normal{1}{0.1} absorb modest under or over dispersions.

For systems with two planets, we reparameterise inclinations as  $(I_1, I_2)\mapsto(I_0, I_{\Delta})$, with
\begin{equation} \label{eq:inclination_cv}
    I_1 = I_0 - I_{\Delta}\,, \qquad   I_2 = I_0 + I_{\Delta}\,. 
\end{equation}

We favour quasi-coplanarity via $I_{\Delta}$$\sim$$\mathcal{N}(0,5.73^\circ)$, which is consistent with multi-planet mutual inclination dispersion \citep{2009ApJ...696.1230F, 2018ApJ...860..101Z, 2020AJ....160..276H}, while allowing separated inclinations if supported by the data. The reference inclination $I_0$\Isotro{0}{180^{\circ}} uses an isotropic prior, uniform in $\cos I_0$ over $[0,1]$.

\section{Target selection and observations}\label{sec:observations}

\begin{table*}[!ht]
\caption{\label{tab:stellar-summary}Stellar parameters summary.}
\centering
    \begin{tabular}{llllll}
    \toprule
    Parameter   &HIP\,8923   &HIP\,10090  &HIP\,21850  &HIP\,39330  &HIP\,98599  \\
    \midrule
    \midrule
    $^1$Spectral type   &G3V   &K0   &G8V   &G5V   &F7V   \\
    $^1$$B-V$           &0.638  &0.766  &0.773  &0.655  &0.556  \\
    $^1$$V$             &8.24   &8.94   &8.41   &7.52   &8.11   \\
    $^1$\smw            &$0.157\pm0.004$ &$0.139\pm0.004$ &$0.135\pm0.004$ &$0.201\pm0.005$ &$0.142\pm0.004$   \\
    $^1$\logrhk         &-5.01   &-5.16   &-5.18   &-4.80   &-5.10   \\
    $^1$\vsini~(\kms)          &3.0   &2.7   &2.8   &3.4   &8.0   \\
    $^1$RV              &$-22.75\pm0.13$   &$9.20\pm0.17$   &$62.63\pm0.12$   &$23.03\pm0.13$   &$-14.78\pm0.13$   \\
    $^2$Mass ($M_{\star}$)            &$1.01^{+0.04}_{-0.04}$  &$0.94^{+0.04}_{-0.04}$   &$^{(3)}$$0.98^{+0.12}_{-0.12}$   &$0.96^{+0.04}_{-0.04}$   &$1.19^{+0.04}_{-0.04}$   \\
    $^{4}$Distance (pc)        &65.87 ± 0.11   &82.30 ± 0.11   &55.50 ± 0.06   &36.39 ± 0.02   &76.10 ± 0.13   \\
    $^{5}$[Fe/H]                &0.42±0.05 &0.47±0.04 &0.43±0.04 &0.16±0.05 &-0.12±0.10 \\
    $^{5}$$T_{\mathrm{eff}}$ (K)&5980±51 &5661±72 &5680±56 &5748±81 &5978±300 \\
    $^{5}$Age (Gyr)             &$1.73_{-1.22}^{+1.14}$ &$4.97_{-1.69}^{+2.87}$ &$3.55_{-2.19}^{+2.48}$ &$6.57_{-1.96}^{+3.00}$ &$5.23_{-3.71}^{+3.12}$ \\
    Time baseline (yr)      &14.80 &16.98 &24.83 &17.01 &20.82 \\
    \bottomrule
    \end{tabular}
    \tablebib{(1)~\citet{1997A&A...323L..49P}; (2)~\citet{2023A&A...669A.104K}, (3)~\citet{2022ApJS..262...21F}, (4)~\citet{2023A&A...674A...1G}, (5) This work with SPECIES~\citet{2018A&A...615A..76S}.}
\end{table*}

Since its inception, CHEPS has targeted quiet, metal-rich FGK stars \citep{2008A&A...485..571J, 2011ASPC..448..991J} with the goal of discovering giant planets \citep{2009MNRAS.398..911J, 2017MNRAS.466..443J}. For the present study we selected CHEPS stars that showed significant long-term RV curvature or acceleration in the 2009-2021 time series and continued monitoring them through 2025, thereby extending individual baselines up to $\sim$16 years. In addition to new CHEPS observations, we assembled all available high-precision archival RVs for these stars from HARPS, CORALIE, UCLES, and HIRES. For this first batch of CHEPS systems, we selected inactive stars (\logrhk$\leq-4.75$), with small \smw~values (\smw$\leq0.201$) which were also slow rotators (\vsini$\leq5$~\kms). Furthermore, we considered targets with a total RV baseline over 14 years and more than 16 CORALIE measurements (see full stellar parameter \reftab{tab:stellar-summary}).

Our RV sample combines high-precision spectra from the High Accuracy Radial Velocity Planet Searcher \citep[HARPS, ][]{2000SPIE.4008..582P} on the ESO 3.6~m telescope at La Silla Observatory; CORALIE \citep{2000fepc.conf..571U} on the 1.2~m Euler telescope at La Silla Observatory; University College London Echelle Spectrograph \citep[UCLES, ][]{1990SPIE.1235..562D} at the Anglo-Australian Telescope (AAT); and the High Resolution Echelle Spectrometer \citep[HIRES, ][]{1994SPIE.2198..362V} at the Keck Observatory. The HARPS data was retrieved from the public HARPS RVBank \citep{2019ascl.soft06004T, 2024A&A...683A.125P}, whilst UCLES and HIRES archival velocities were gathered from published releases and institutional archives.

HARPS experienced a fibre upgrade (2015-06-02) and a warm-up (2020-03-23) post-COVID shutdown \citep{2015Msngr.162....9L}. Therefore, we treat HARPS with three different offsets: H (pre-2015), H15 (2015-2020), and H20 (post-2020).

CORALIE underwent a major upgrade in 2014, the circular fibre link was replaced with octagonal fibres, and a Fabry-Pérot calibration unit was added \citep{2019A&A...625A..71R}. We adopt three subsets consistent with CHEPS processing: COR07 (pre-upgrade), COR14T (post-upgrades, reduced with TERRA \citet{2012ApJS..200...15A}), and COR14 (post-upgrades, reduced with the CORALIE DRS).

We use the Hipparcos IAD from the new 2007 reduction \citep{1997A&A...323L..49P, 2007A&A...474..653V}, obtaining the data through \texttt{htof}\,\citep{2021ApJS..254...42B}, and Gaia DR3 five-parameter solutions \citep{2023A&A...674A...1G}. To project our orbital models into Gaia's measurement geometry at the Gaia reference epoch, we query GOST for scan angles.

A summary table with the number of measurements per target can be found in \reftab{tab:n-obs-target}.

\section{Analysis and results}\label{sec:results}
Within the \emp\, framework, joint astrometry is straight-forward to use, and a step-by-step tutorial can be found in the online documentation\footnote{\scriptsize\url{https://astroemperor.readthedocs.io/en/latest/tutorials/astrometry}}.

For each target, we examine their RVs with Generalised Lomb-Scargle periodograms \citep[GLS, ][]{1976Ap&SS..39..447L, 1982ApJ...263..835S}, as well as with correlograms, calculating the Pearson-correlation coefficients between the RVs and stellar activity indices. We then begin fitting models with \emp, first by testing the null hypothesis that comprises just a mean and white noise (hereinafter \zerok). After this, we incrementally add Keplerian signals (models dubbed \onek, \twok, and so on), and compare the Bayesian evidence \lnzh~with the previous model; with a difference of $\Delta$\lnzh$\geq5$ we select the new model. Each model is run 7 different times, to ensure we have consistent results. The values we present for model comparison and parameter estimation correspond to the run with the median evidence. Hyper-parameter selection for our runs was set to 20 temperatures, 256 walkers, 80\,000 sweeps, and 1 step, for a grand total of 409\,600\,000 samples. Ladder adaptation is done with a Uniform Swap Acceptance Rate kernel, with rate $\nu_0=1.0$, and timescale $\tau_0=100$. We leave the adaptation for 40\,000 sweeps, and then freeze the ladder. This initial phase we consider as burn-in, so these samples are dropped before any statistical inferences are made, preserving detailed balance so that standard ergodicity theorems apply.

As a validation case, we first analyse HIP\,21850, a well-studied system hosting two confirmed exoplanets. We take as references the parameter solutions of \citet[][RV-only]{2017AJ....154..274W} and \citet[][joint RV+astrometry]{2022ApJS..262...21F}, hereinafter W17 and F22, respectively. A summary of planet signatures for all systems can be found in \reftab{tab:planets-summary}.

\begin{table*}[ht!]
\centering
\caption{\label{tab:planets-summary}Planet signatures summary for joint RV+astrometry fit}, with Jupiter analogues in bold.
\begin{tabular}{llllll}
\toprule
Planet    &Period (days) &Mass ($M_J$) & a (AU) & e & I ($^{\circ}$)\\
\midrule
\midrule
\textbf{HIP\,8923b}   &$5160^{+150}_{-240}$ &$9.98_{-0.16}^{+0.78}$ &$5.90_{-0.22}^{+0.10}$ &$0.011^{+0.032}_{-0.011}$ &$24.8^{+1.2}_{-3.9}$ \\
\textbf{HIP\,10090b}  &$2960^{+120}_{-100}$ &$3.87_{-0.60}^{+0.65}$ &$3.87_{-0.11}^{+0.19}$ &$0.115^{+0.005}_{-0.033}$ &$156.5^{+1.8}_{-2.2}$\\
HIP\,10090c           &$321.8^{+0.3}_{-0.6}$&$0.85_{-0.12}^{+0.03}$ &$0.90_{-0.01}^{+0.01}$ &$0.188^{+0.043}_{-0.048}$ &$153.8^{+1.3}_{-3.1}$\\
\textbf{HIP\,21850b}  &$2539^{+1}_{-3}    $ &$8.25_{-0.85}^{+1.05}$ &$3.49_{-0.14}^{+0.18}$ &$0.186^{+0.002}_{-0.004}$ &$65.5^{+1.7}_{-4.2}$ \\
HIP\,21850c           &$11320^{+490}_{-940}$&$4.67_{-0.43}^{+0.44}$ &$9.90_{-0.62}^{+0.29}$ &$0.210^{+0.010}_{-0.020}$ &$72.7_{-5.8}^{+17.8}$\\
\textbf{HIP\,39330b}  &$4650^{+210}_{-250} $&$1.67_{-0.08}^{+0.19}$ &$5.05_{-0.17}^{+0.23}$ &$0.078^{+0.053}_{-0.078}$ &$55.0^{+0.6}_{-3.9}$\\

\textbf{HIP\,98599b}  &$2656^{+40}_{-16}  $&$6.85_{-0.22}^{+0.10}$ &$3.96_{-0.01}^{+0.07}$ &$0.169^{+0.002}_{-0.055}$ &$29.7^{+6.9}_{-6.2}$ \\
\bottomrule
\end{tabular}
\end{table*}

\subsection{HIP 21850}\label{sec:res_hip21850}

\begin{figure} 
    \includegraphics[width=\columnwidth]{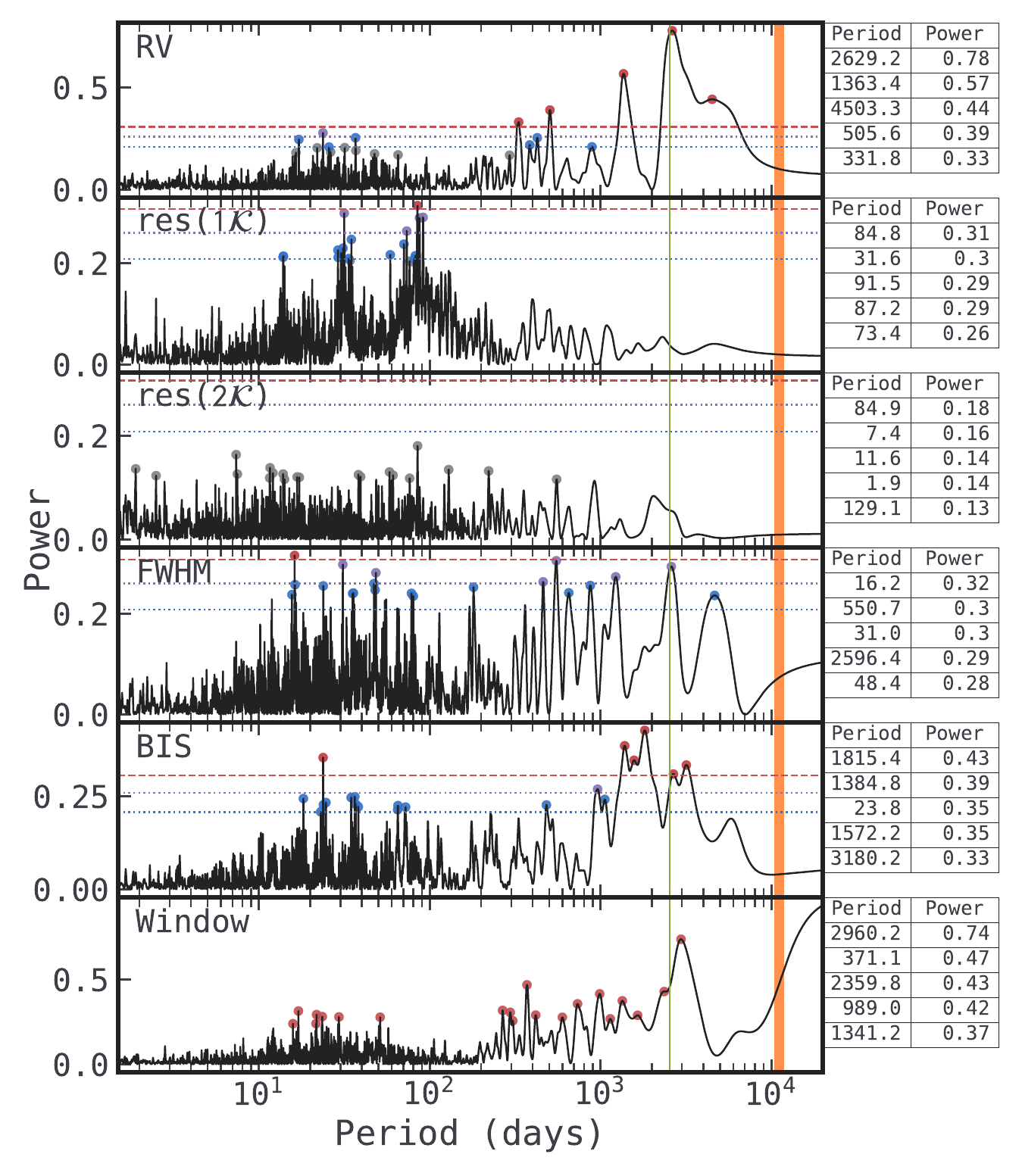}
    \caption{\label{fig:hip21850-periodogram}HIP\,21850 periodogram. Descending, RVs, model residuals, FWHM, BIS, and the window function. FAP lines included for 0.1, 1 and 10\%, in dashed red, dotted purple, and dotted blue, respectively. Circle markers show the periods with greatest power, coloured by FAP region. The coloured regions correspond to $P_1=2539^{+0.8}_{-3.1}$~d (green) and $P_2=11320^{+490}_{-940}$~d (orange).}
\end{figure}

\begin{figure} 
    \includegraphics[width=\columnwidth]{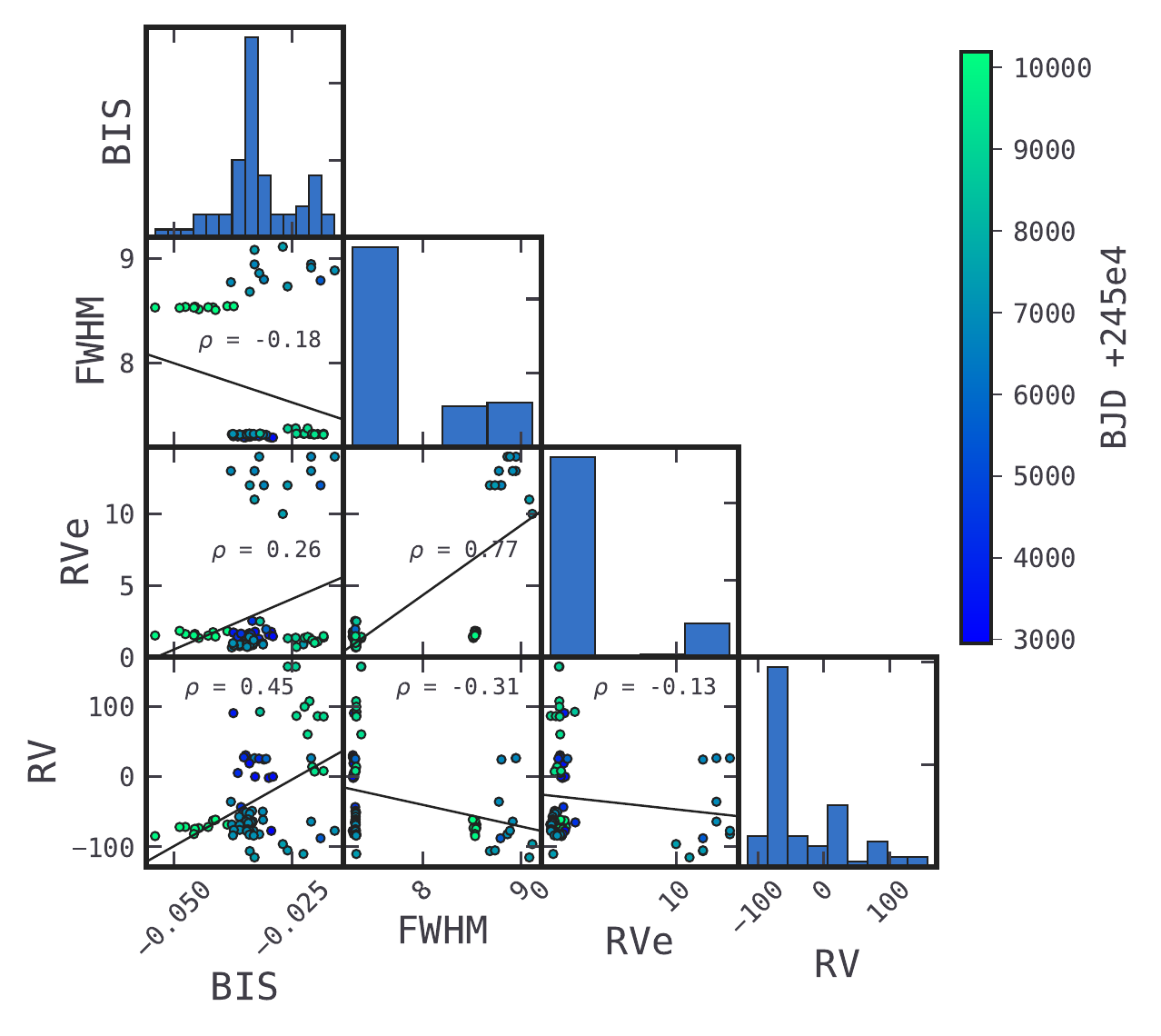}
    \caption{\label{fig:hip21850-correlogram}HIP\,21850 correlogram. Each panel displays the pairwise relation between RVs, full-width half-maximum (FWHM), CCF bisector inverse slope (BIS), and formal RV uncertainty (RVe).  Pearson correlation coefficient is denoted by $\rho$, with the linear trend as a black line.}
\end{figure}

Also known as HD\,30177, in the GLS periodogram, the RVs present a highly significant peak--above the 99.9\% false-alarm probability (FAP) line--at 2629~d. The GLS of the full-width half-maximum (FWHM) has a marginally significant peak--between the 90\% and 99.0\% FAP lines--at 2596~d (see \reffig{fig:hip21850-periodogram}), while the bisector inverse slope (BIS) presents a highly significant peak at 2671~d. The window function presents a dominant peak at 2960~d. On the other hand, the correlogram reveals weak to moderate correlations ($|\rho| \in [0.20,0.60[$) between RVs and indices: $\rho_{\mathrm{RV, FW}}$=-0.31 for FWHM and $\rho_{\mathrm{RV, BIS}}$=0.45 for BIS (see \reffig{fig:hip21850-correlogram}).

With the \emp~framework we start by analysing RVs alone. The single Keplerian solution (\onek) finds an orbital period of $P_1$=$2534^{+22}_{-11}$~d, and adding a second Keplerian (\twok) yields periods $P_1$=$2539.8^{+4.1}_{-4.7}$~d and $P_2$=$15900^{+2500}_{-4800}$~d. Model comparison is decisive for the \twok\,model, as the Bayes' factor $\Delta$\lnzh=$102.4$ strongly favours it, with a Bayesian evidence of \lnzh$\mid_{2\mathcal{K}}$=$-389.5\pm0.3$ over the \onek\, model's \lnzh$\mid_{1\mathcal{K}}$=$-491.9\pm0.2$. Each planet has minimum mass and semi-major axis estimates of \msini$_1=8.2^{+0.1}_{-0.5}~M_J$, \msini$_2=6.6^{+1.4}_{-2.5}~M_J$, $a_1=3.65_{-0.11}^{+0.02}$, and $a_2=12.5_{-2.7}^{+1.4}$.

Running \emp~with RV+astrometry keeps the preference the same, \lnzh$\mid_{2\mathcal{K}}$=$-713.4\pm0.3$ over \lnzh$\mid_{1\mathcal{K}}$=$-817.5\pm0.3$, with a Bayes' factor of $\Delta$\lnzh=$104.10$ (extensive discussion on Bayes' factor differences with the addition of astrometry in \refsec{sec:discussion-model-evidence}). This time, we find $P_1$=$2539^{+1}_{-3}$~d and $P_2$=$11320^{+490}_{-940}$~d, with true masses and semi-major axes $M_1=8.25_{-0.85}^{+1.05}~M_J$, $M_2=4.67_{-0.43}^{+0.44}~M_J$, $a_1=3.49_{-0.14}^{+0.18}$, and $a_2=9.90_{-0.62}^{+0.29}$ (see \reftab{tab:hip21850_params}).

\begin{figure} 
    \includegraphics[width=\columnwidth]{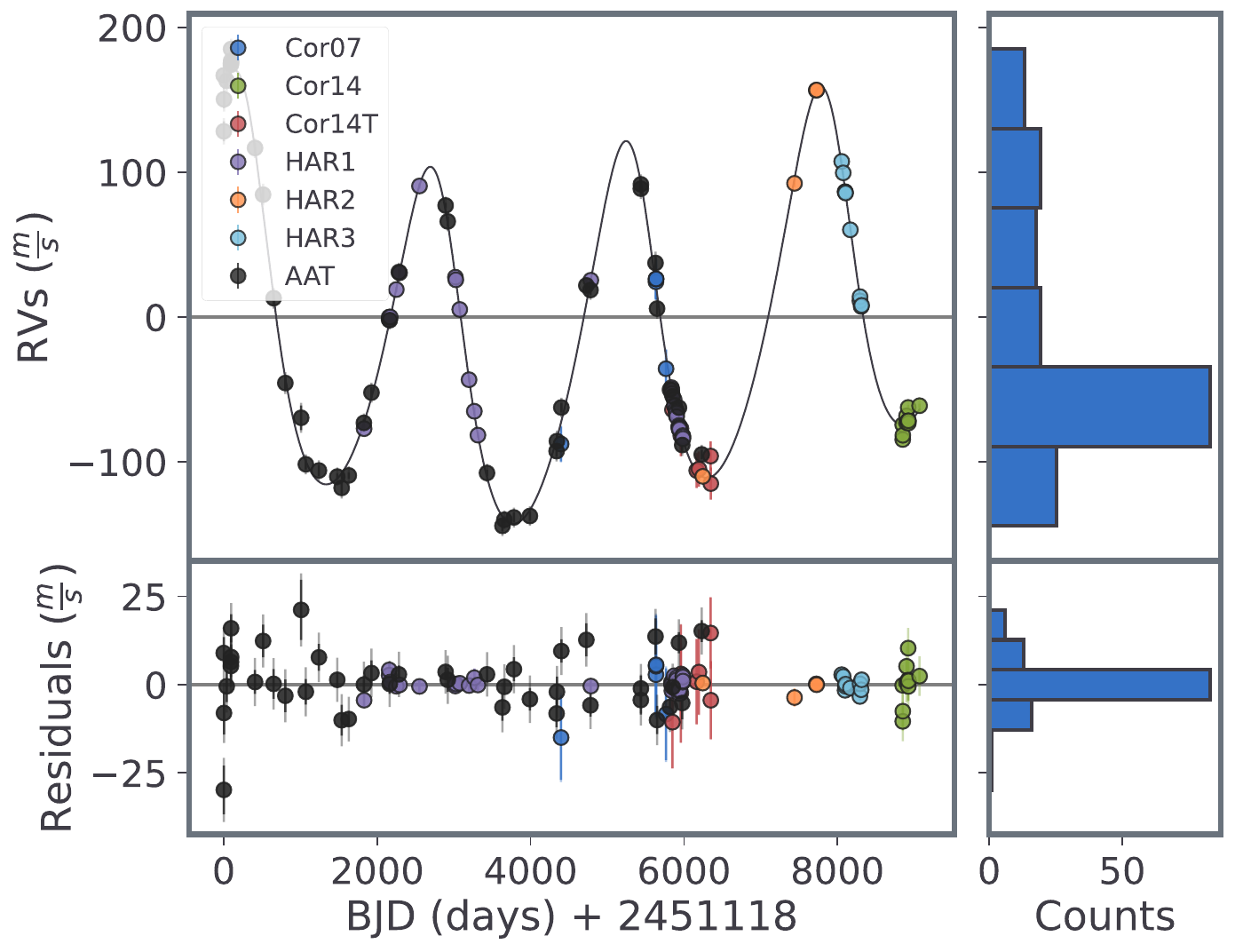} \caption{HIP\,21850 RVs as circles coloured per instrument with the \twok~model imposed as a black line. At the bottom, RV residuals, and at the right of each plot, RV distribution histograms.}\label{fig:hip21850-keplerian}

    \includegraphics[width=\columnwidth]{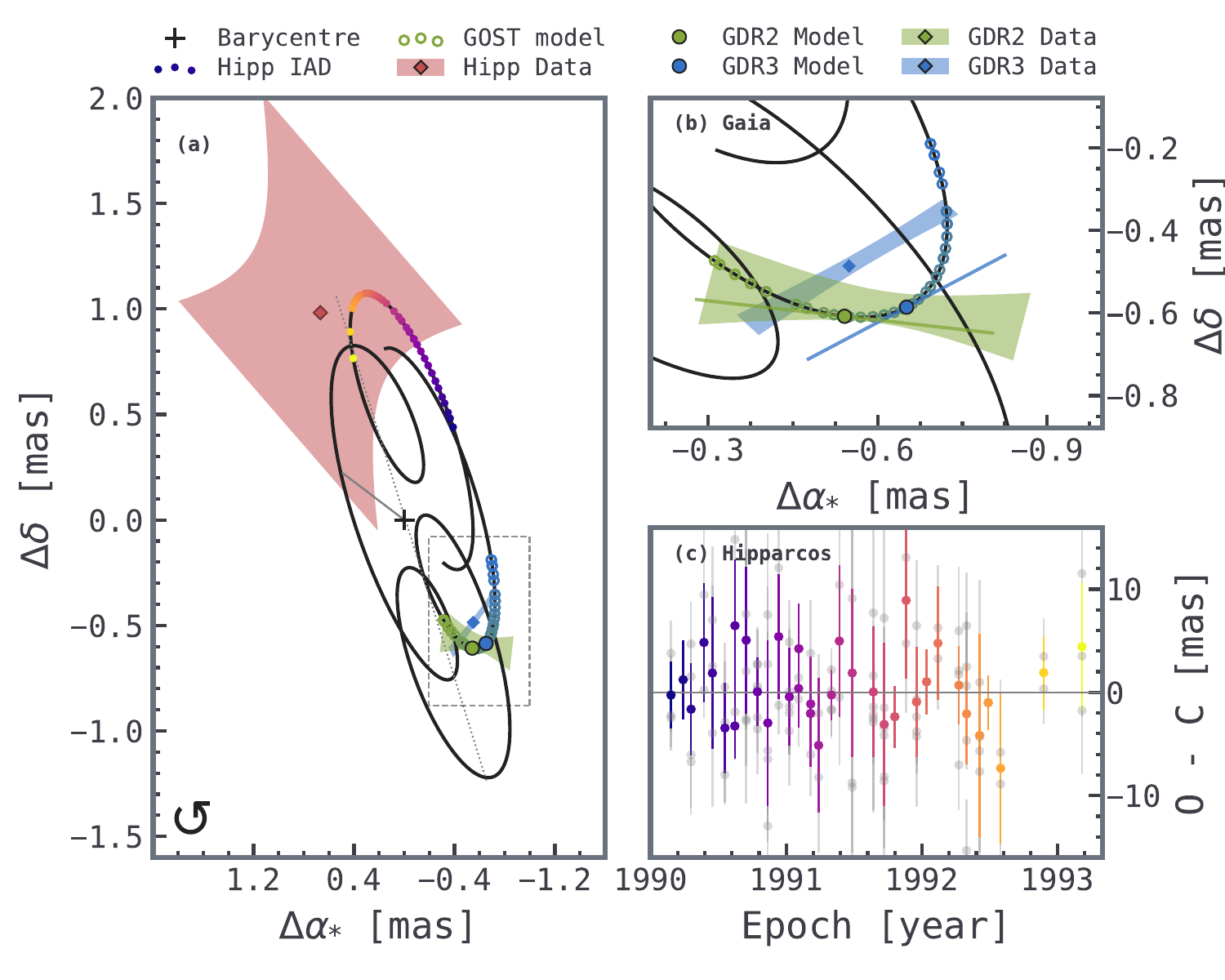}
    \caption{HIP\,21850 astrometry best-fit for the \twok~model. Displaying (a) the best-fit astrometric orbit, (b) zoom-in for Gaia and GOST model, gray rectangle in panel (a), and (c) residuals for Hipparcos abscissa.}\label{fig:hip21850-am-plot}
\end{figure}

The inclusion of astrometry helps significantly constraining the second period, $P_2$=$15900^{+2500}_{-4800} \rightarrow 11320^{+490}_{-940}$~d, while determining the true masses. The posteriors for the \twok~RV-only model reveal that $P_2$ sits uncomfortably in a posterior with a flattened top between $\sim$9~000--17~000~d, attested by the loose uncertainties. The RV+astrometry posteriors round significantly, still with a flattened top around 11\,000--12\,000~d, which adds a clear upper boundary on the solution. Our solution remains fairly similar to W17 and F22, considering that with the addition of the CHEPS RV data, we extend the baseline from $\sim$6\,000~d to 9\,070~d. The RV part of the joint-model fit can be seen in \reffig{fig:hip21850-keplerian}. And the astrometry in \reffig{fig:hip21850-am-plot}. Further discussion on parameter estimates can be found in \refsec{sec:discussion-improved-planet-parameters}.

For consistency we run the RV residuals through the GLS once again: the RV field is flattened, with a single marginally significant peak 90\% FAP line at 91.1~d. Similarly, the residuals correlogram is flat, with $\rho_{\mathrm{RV, FW}}$=0.01 and $\rho_{\mathrm{RV, BIS}}$=-0.07. 
The flat residuals periodogram and correlogram, plus the strong Bayesian preference, and internal consistency of the $P_1$ solution, are good indicators for a planetary companion: if stellar activity were the origin of the main RV peaks, we would expect coherent residual power and/or correlations. Furthermore, all RV signals discussed in this work have larger amplitudes than what would be expected to be produced by stellar activity for these stars \citep[up to a few \ms, ][]{2009ApJ...707L..73M, 2010A&A...512A..39M}. Further stellar activity discussion and modelling with GPs with the \emp~framework is extensively discussed in its paper \cite{emperor-paper}.

Finally, is worth discussing that the window function, initially a plausible source for some GLS structure given the peaks at 2960 and 2360~d, is unlikely to shape this solution: For the \onek~the fitted $P_1=2539$~d signal sits between both window function spikes, and survives in between when moving onto \twok, and even when including astrometry to the fit.

\begin{table*}
\caption{\label{tab:hip21850_params}HIP\,21850 joint RV+astrometry parameter estimates.}
\centering
\begin{tabular}{llllll}
\toprule
   Parameter      & Prior             & \citet{2017AJ....154..274W} & \citet{2022ApJS..262...21F} & This work \\ 
\midrule
\midrule

$P_1$~(days)      & \Uniform{1}{20000}& 2524.4±9.8  &$2514.5^{+5.2}_{-4.3}$    &$2539^{+1}_{-3}$    \\ 
$K_1$~(\ms)       & \Uniform{0+}{200} & 126.3±1.5   &$128.9^{+2.4}_{-2.2}$     &$124.1^{+1.4}_{-1.0}$     \\ 
$e_1$             & \Normal{0}{0.3}   & 0.184±0.012 &$0.207^{+0.012}_{-0.017}$ &$0.186^{+0.002}_{-0.004}$ \\ 
$I_1$~$^{\circ}$  & \Isotro{0}{180}   & -           &$85.39^{+14.35}_{-18.74}$ &$65.5^{+1.7}_{-4.2}$   \\ 
$a_1$~AU          & -                 & 3.58±0.01   &$3.60^{+0.14}_{-0.15}$    &$3.49_{-0.14}^{+0.18}$    \\ 
$M_{\rm p1}$~$M_J$& -                 & -           &$8.40^{+1.24}_{-0.49}$    &$8.25_{-0.85}^{+1.05}$    \\ 

\midrule

$P_2$~(days)      & \Uniform{1}{20000}&11613±1837   &$12085^{+583}_{-441}$     &   $11320^{+490}_{-940}$ \\ 
$K_2$~(\ms)       & \Uniform{0+}{200} &70.80±29.50  &$54.47^{+5.17}_{-3.99}$   & $41.03^{+1.60}_{-2.93}$ \\ 
$e_2$             & \Normal{0}{0.3}   &0.22±0.14    &$0.04^{+0.01}_{-0.01}$    &  $0.21^{+0.01}_{-0.02}$ \\ 
$I_2$~$^{\circ}$  & \Isotro{0}{180}   &-            &$98.02^{+16.03}_{-24.24}$ &$72.7_{-5.8}^{+17.8}$ \\ 
$a_2$~AU          & -                 &9.890±1.040  &$10.26^{+0.55}_{-0.48}$   & $9.90_{-0.62}^{+0.29}$  \\ 
$M_{\rm p2}$~$M_J$& -                 &-            &$6.15^{+1.31}_{-0.34}$     & $4.67_{-0.43}^{+0.44}$  \\ 

\bottomrule
\end{tabular}
\end{table*}

\subsection{HIP 8923}\label{sec:res_hip8923}

Also known as HD\,11731, the GLS periodogram (see \reffig{fig:hip8923-periodogram}) shows three highly significant peaks (>99.9\% FAP line) in the RVs, at 5020~d, 314~d--corresponding with the yearly alias between 5020~d and 337~d in the window function--and 29.7~d--the moon cycle, also present in the window function. The correlogram, on the other hand, shows a strong correlation (here defined as $|\rho| \in[0.60, 0.80[$) in $\rho_{\mathrm{RV, FW}}$=-0.63. This correlation might be an artifact due to the uncertainty difference between datasets, since there is a very strong correlation ($\rho \in[0.80, 1[$) between RV uncertainty and FWHM $\rho_{\mathrm{RVe, FW}}$=0.97.

Our RV-only search yielded a single planet with orbital period $P_1=5120_{-260}^{+460}$~d, with evidence \lnzh$\mid_{1\mathcal{K}}=-87.64\pm0.11$, and a Bayes' factor of $\Delta$\lnzh=$16.77$ over the white-noise-only model \lnzh$\mid_{0\mathcal{K}}=-104.41\pm0.09$. By adding the astrometry data, we find 
$P_1=5160^{+150}_{-240}$~d, consistent with the RV-only model. In this scenario, the evidence is \lnzh$\mid_{1\mathcal{K}}=-297.97\pm0.19$, with a $\Delta$\lnzh=75.35 over the \zerok~model \lnzh$\mid_{0\mathcal{K}}=-373.32\pm0.23$.

\begin{figure} 
    \includegraphics[width=\columnwidth]{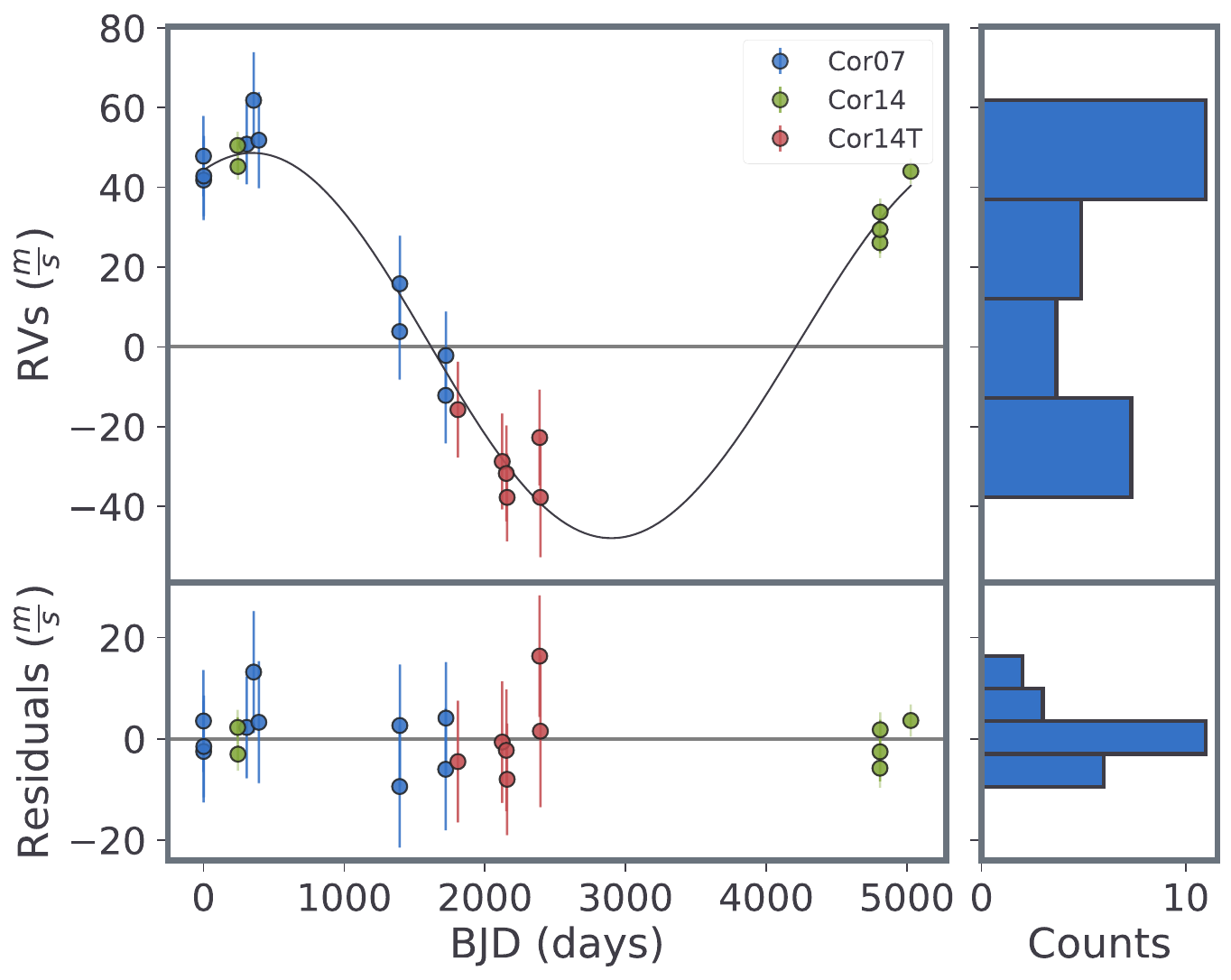}
    \caption{HIP\,8923 RVs phase-folded at $P=5160^{+150}_{-240}$~d as circles coloured per instrument with the \onek~model imposed as a black line. Bottom, RV residuals. Right, RVs histograms.}
    \label{fig:hip8923-keplerian}

    \includegraphics[width=\columnwidth]{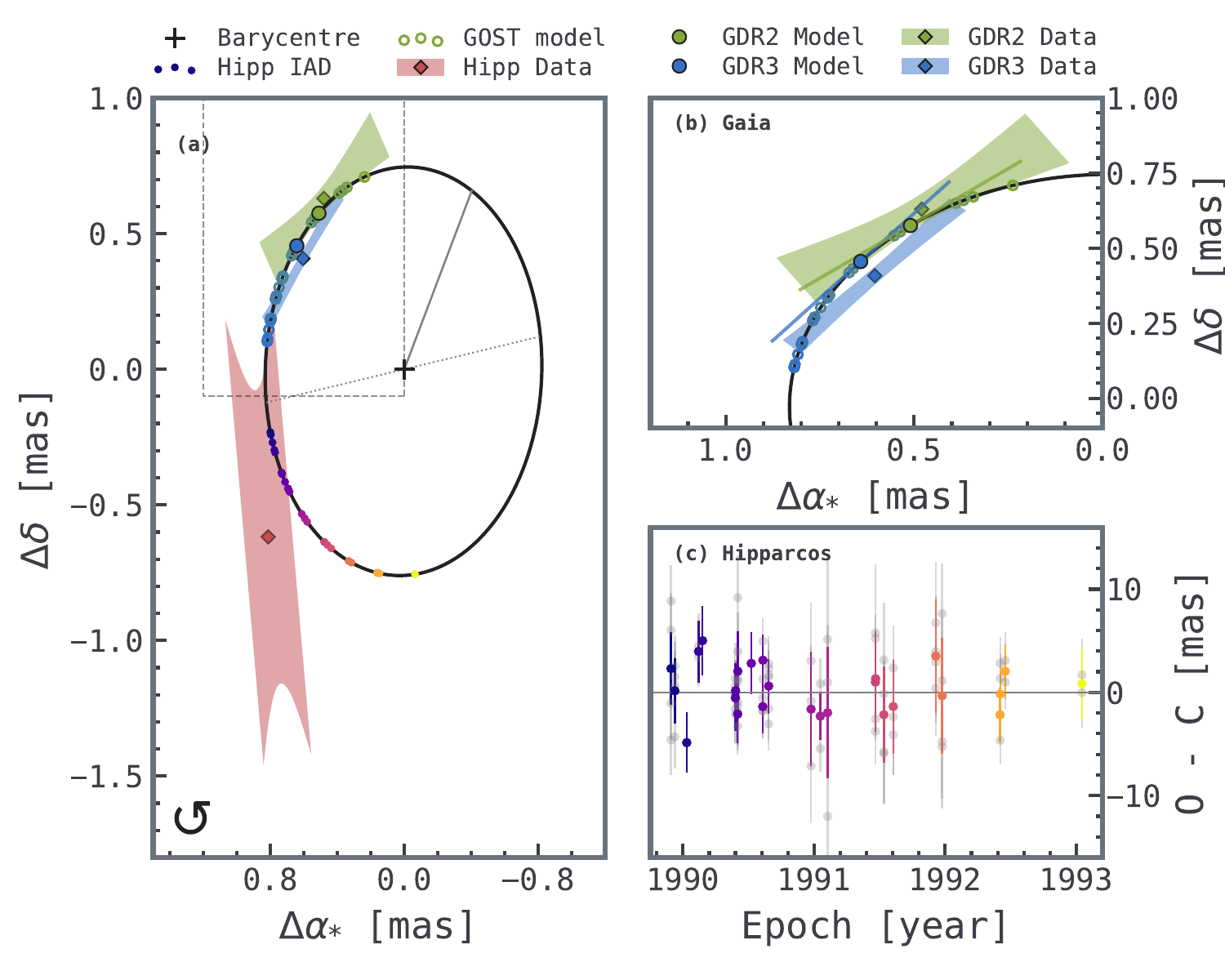}
    \caption{HIP\,8923 best-fit for the \onek~model. Displaying for panel (a) the best-fit astrometric orbit, (b) zoom-in for Gaia and GOST model, gray rectangle in panel (a), and (c) residuals for Hipparcos abscissa.}
    \label{fig:hip8923-am-plot}
\end{figure}

This case strongly demonstrates the value of astrometry when only a part of a long-period orbit is sampled by RVs (see \reffig{fig:hip8923-keplerian}). The relative Bayes' factor between \zerok~and \onek~increases significantly when adding astrometry, $\Delta$\lnzh=$16.77\rightarrow75.35$. Furthermore, the inclusion of the astrometric data shrunk the period uncertainty almost by half $P_1$=$5120_{-260}^{+460} \mapsto 5160^{+150}_{-240}$~d, where the relative uncertainty corresponds to 13.97\%, and 7.57\% respectively. Similarly, the minimum mass \msini=$4.30_{-0.75}^{+0.07}~M_J$ becomes an absolute mass of $M_1=9.98_{-0.16}^{+0.78}~M_J$, and semi-major axis $a_1=5.89^{+0.30}_{-0.26} \mapsto 5.90^{+0.10}_{-0.22}$. The complete astrometric fit can be seen in panel (a) of \reffig{fig:hip8923-am-plot}, where the black cross is the system's barycentre, the gray line connects it to the periapsis, and the gray dotted line joins ascending and descending nodes. Reference epochs are shown as diamonds with their position and proper motion uncertainty as hourglass-shaped shaded regions. Solid circles and slopes show the best-fit position and proper motion induced by the companion. Green to blue coloured rings show GOST data projected with the best-fit. Panel (b) is a zoom-in of the region comprised of GDR2, GDR3, and GOST, delimited in panel (a) with a gray dashed box. Panel (c) shows the residuals of Hipparcos abscissa as solid grey circles, where multiple measurements per epoch have been binned and coloured by time.

Finally, we address the peak that appears in the window function at  4818~d. Even though it is ranked 12th in height, we analyse the residuals of the model: the GLS leaves no significant peaks above the 90\% FAP line, strongly arguing that the ~5\,000-day power is well explained by the orbit rather than harmonics or aliases. On the other hand, the residuals' correlogram shows $\rho_{\mathrm{res, FW}}$=0.28, a significant shift from the initial $\rho_{\mathrm{RV, FW}}$=-0.63. Even though this correlation is most likely an artifact due RV uncertainties between datasets, we observe that the strong anti-correlation disappears once the Keplerian is removed, which is consistent with the planet driving most of the long-term RV variation, and not stellar activity. We also note that the yearly alias producing the second highest peak in the RVs at 314~d, completely disappears in the residuals.

\subsection{HIP 10090}\label{sec:res_hip10090}
For HIP\,10090, also known as HD\,13350, our GLS periodogram reveals a dominant highly significant peak at 2892~d, another at 324~d, and a family of peaks around 25-32~d, corresponding to the lunar cycle. Stellar activity indices don't show any peaks above the 90\% FAP line, and the window function shows peaks at 364 and 29.7~d, corresponding to the yearly and monthly aliases. The correlogram shows no significative correlations with stellar activities ($|\rho|<0.2$).

With \emp, the RV-only solution settles at \twok\,with period $P_1=2980^{+140}_{-100}$~d and $P_2=321.3^{+0.5}_{-0.4}$~d, with evidence \lnzh$\mid_{2\mathcal{K}}=-203.84\pm0.18$, and a Bayes' factor of 10.92 over the \lnzh$\mid_{1\mathcal{K}}$. The runs including astrometry data settle at \twok\,as well, with $P_1=2960^{+120}_{-100}$~d and $P_2=321.8^{+0.3}_{-0.6}$~d, consistent with the RV-only solution. This run has \lnzh$\mid_{2\mathcal{K}}=-548.57\pm0.19$, with a Bayes' factor of 27.22 over \lnzh$\mid_{1\mathcal{K}}$. By analysing the residuals of this model, the GLS stays flat, with no peaks over the 90\% FAP line, while correlations remain not significant. Once again, astrometry drives to a precise mass estimation \msini$_1=1.53_{-0.14}^{+0.12} \mapsto M_1=3.87_{-0.60}~M_J$, and \msini$_2=0.39_{-0.03}^{+0.02} \mapsto M_2=0.85_{-0.12}^{+0.03}~M_J$.

\begin{figure} 
    \includegraphics[width=\columnwidth]{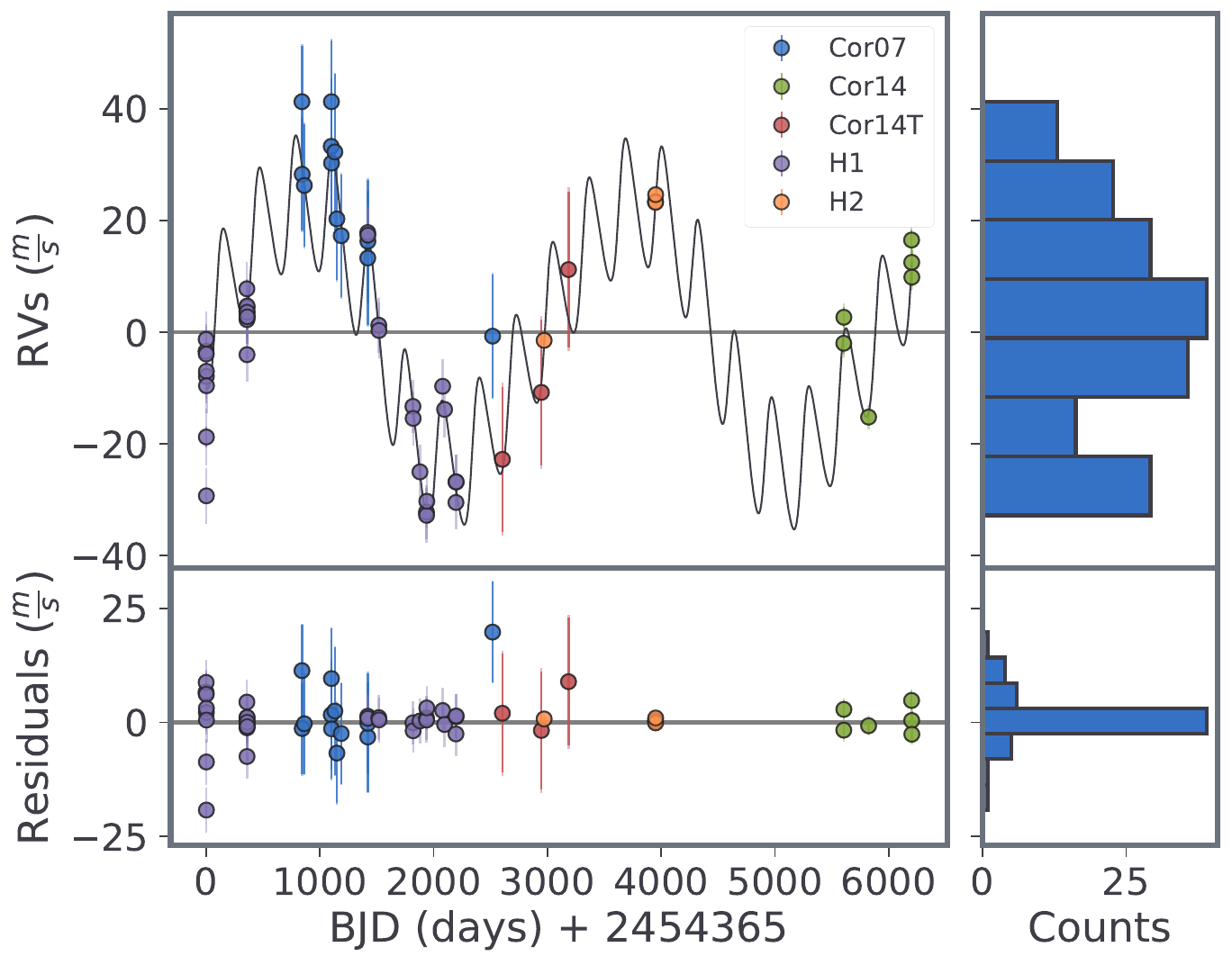}
    \caption{HIP\,10090 RVs as circles coloured per instrument with the \twok~model imposed as a black line. Bottom, RV residuals. Right of each plot, RVs histograms.}
    \label{fig:hip10090-keplerian}

    \includegraphics[width=\columnwidth]{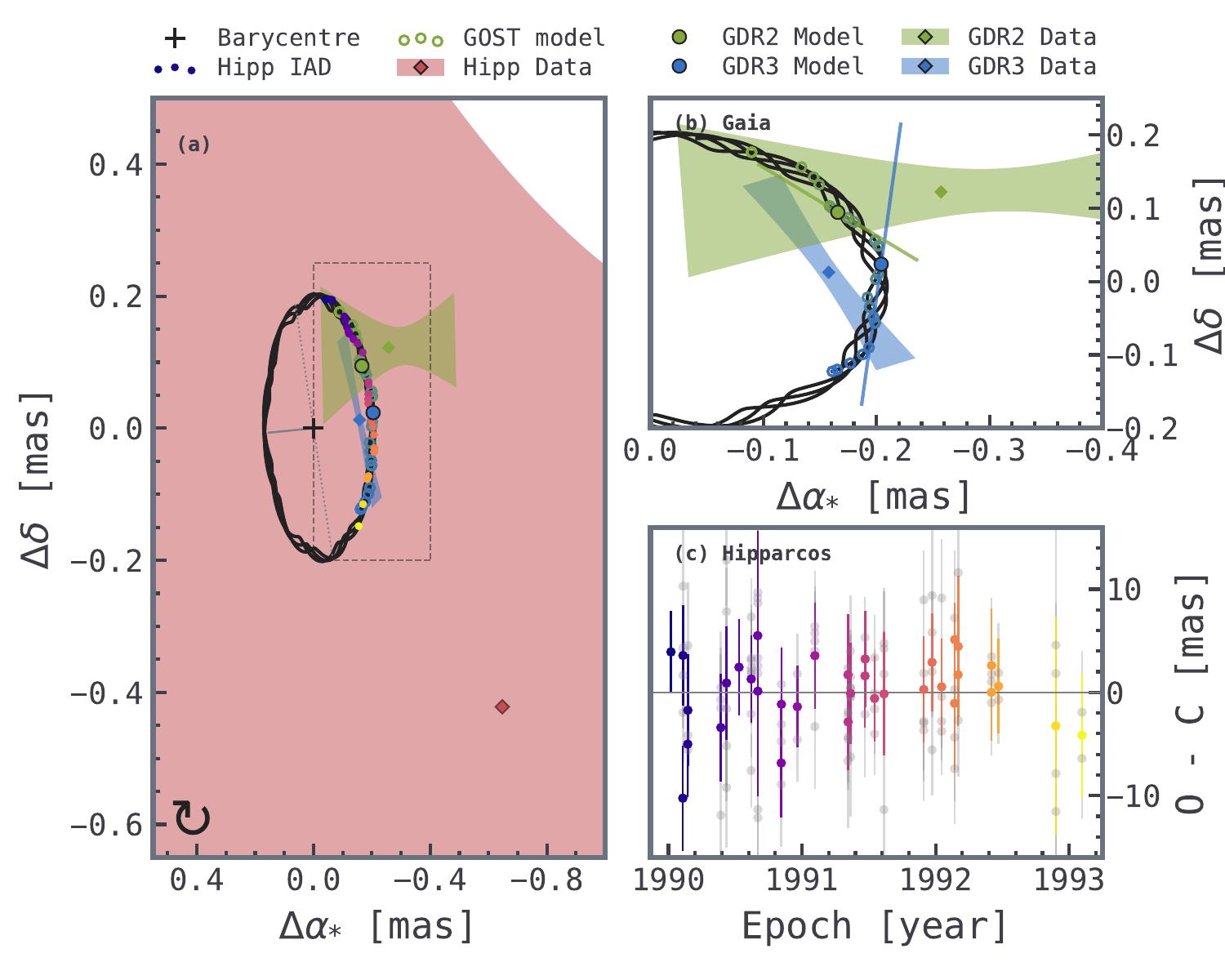}
    \caption{HIP\,10090 best-fit for the \twok~model. Displaying for panel (a) the best-fit astrometric orbit, (b) zoom-in for Gaia and GOST model, gray rectangle in panel (a), and (c) residuals for Hipparcos abscissa.}
    \label{fig:hip10090-plot_am}
\end{figure}

Hipparcos relatively high proper motion (red ellipse in \reffig{fig:hip10090-plot_am}) makes this part of the model relatively uninformative compared to Gaia GDR2 and 3. We further discuss this point in \refsec{sec:discussion-model-evidence}.

\subsection{HIP 39330}\label{sec:res_hip39330}

For HIP\,39330, also known as HD\,66653, the GLS periodogram shows a single dominant peak in the RVs (see \reffig{fig:hip39330-periodogram}) 4547~d. The FWHM periodogram has significant peaks at 17.9~d and 29.0~d, the latter corresponding with the moon cycle. The correlogram does not reveal significant correlations ($|\rho|<0.2$) between RVs and stellar activities (see \reffig{fig:hip39330-correlogram}). 

The RV-only solution settles at \onek~with $P_1=4710_{-150}^{+20}$~d, with \lnzh$\mid_{1\mathcal{K}}$=$-202.41\pm0.14$. The \twok~solution presents higher evidence \lnzh$\mid_{2\mathcal{K}}=-198.48\pm0.17$, but it does not comply with our previous requirement of $\Delta$\lnzh$=3.93<5$, finding orbital periods $P_1=4640^{+300}_{-200}$~d, and $P_2=257^{+160}_{-75}$~d. 

\begin{figure} 
    \includegraphics[width=\columnwidth]{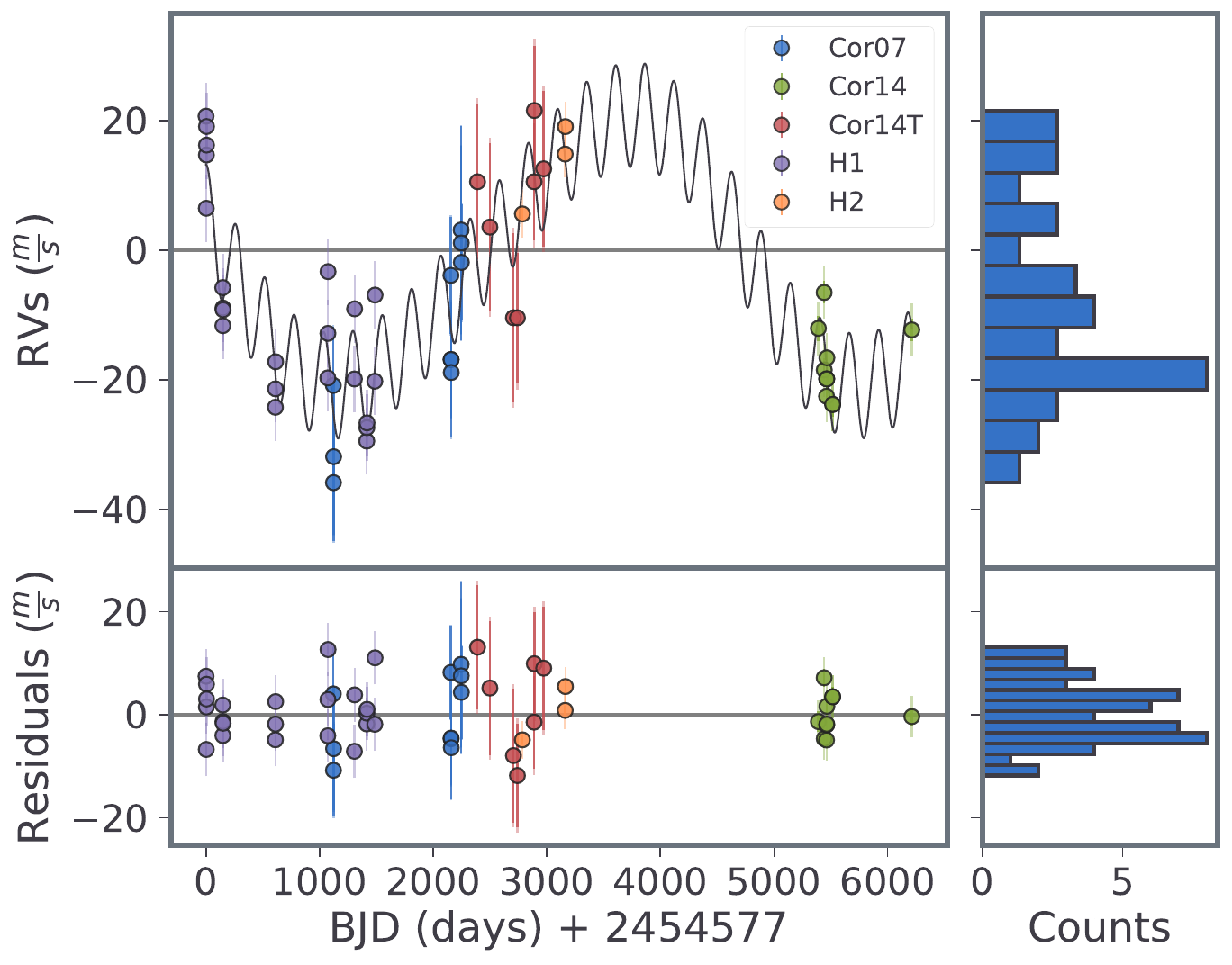}
    \caption{HIP\,39330 RVs as circles coloured per instrument with the \twok~model imposed as a black line. At the bottom, RV residuals, and at the right of each plot, RV distribution histograms.}
    \label{fig:hip39330-keplerian}

    \includegraphics[width=\columnwidth]{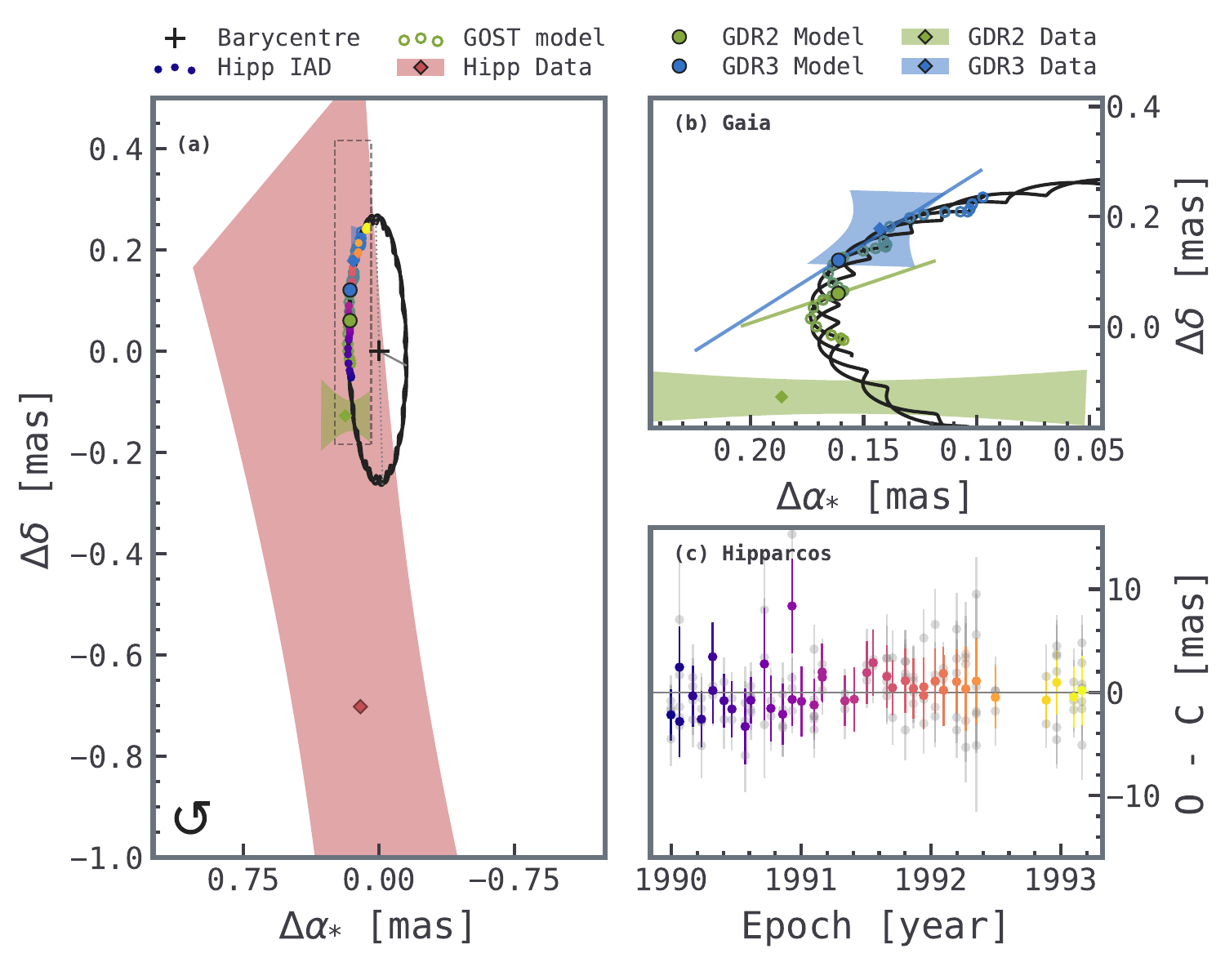}
    \caption{HIP\,39330 best-fit for the \twok$_1$~model. Displaying for panel (a) the best-fit astrometric orbit, (b) zoom-in for Gaia and GOST model, gray rectangle in panel (a), and (c) residuals for Hipparcos abscissa.}
    \label{fig:hip39330-plot_am}
\end{figure}

The joint RV+astrometry analysis settles at \twok~with \lnzh$\mid_{2}=-548.83\pm0.23$, with a $\Delta$\lnzh$=6.97$ over the \onek, with $P_1=4596.9^{+1.7}_{-154.0}$~d, and $P_2=257.1^{+1.6}_{-1.4}$~d. Out of our sample, this is the only system that presents equally likely multiple peaks (see \reffig{fig:hip39330-cornerplot}). 
Our sampler, or any MCMC sampler, is characterised by not returning point estimates for each parameter, but full distributions. And since we are modelling an astrophysical phenomenon that in reality must have a single solution (and quantum orbits are not fashionable) we try to identify a unique orbital configuration by treating model uncertainty and handling it explicitly. \citet{2024MNRAS.534.2858X} assesses that for astrometry data, two data points (i.e., Hipparcos and GDR3) are insufficient to determine inclination, resulting in a bi-modal solution. A third point--from GDR2--resolves this conundrum for their orbital fitting. We diagnose this phenomenon as the origin of our multimodality. For this particular system, GDR2 leaves little information (poor fit, see \reffig{fig:hip39330-plot_am}), leaving us with a double headed dragon.


Consequently, we try fitting modes separately and quantifying the odds. Instead of slicing the parameter space arbitrarily, we set a small normal prior on each possible inclination, with \twok$_1$\Normal{0.92}{0.3}, and \twok$_2$\Normal{2.29}{0.3}. This results in $\Delta$\lnzh=0.15 marginally favouring \twok$_2$ (parameter summary for both models in \reftab{tab:HIP39330_params}). We take the \onek~model as our solution, and further discuss this system in \refsec{sec:discussion}. We also take note of the third and second highest RV-periodogram peaks at 403.8~d (possibly corresponding to the yearly alias with $P_1$) and 57.7~d, which subsequently disappear in the residuals of \onek~and \twok.

\subsection{HIP 98599}\label{sec:res_hip98599}

For HIP\,98599 (or HD\,189627), the GLS (see \reffig{fig:hip98599-periodogram}) reveals a single peak dominating the RVs at 2695~d, and the FWHM GLS has highly significant peaks surrounding it, at 1819.0~d, and 3953.9~d. We see a similar structure to that of the FWHM in the BIS, although with no significant power. The correlogram shows moderate to strong correlations ($|\rho|\in[0.4, 0.8[$) between the RVs with both FWHM and BIS: $\rho_{\mathrm{RV, FW}}$=0.64, $\rho_{\mathrm{RV, BIS}}$=0.51. Similarly to HIP\,8923, both correlations might be artifacts due the uncertainty differences between datasets: correlations between RV uncertainty and indices are strong to very strong $\rho_{\mathrm{RVe, FW}}$=0.96, and $\rho_{\mathrm{RVe, BIS}}$=0.62. This is further sustained by examining correlations in HARPS datasets alone, $\rho_{\mathrm{RV_{HAR}, FW}}$=-0.26, and $\rho_{\mathrm{RV_{HAR}, BIS}}$=-0.11, shifting from strongly positive to weakly negative.

The \emp~RV-only analysis settles at \onek~with \lnzh$\mid_{1\mathcal{K}}=-144.09\pm0.14$, with $\Delta$\lnzh$=47.50$ over the \zerok, with $P_1=2580^{+80}_{-90}$~d. The RV+astrometry analysis settles at \onek~as well, with $P_1=2656^{+40}_{-16}$~d, with \lnzh$\mid_{1\mathcal{K}}=-543.01\pm0.20$, and $\Delta$\lnzh$=61.12$ over the \zerok~model, a Bayes' factor difference of 13.62 with the RV-only model. This result also matches the main GLS peak in the RVs. Furthermore, examining the residuals of the GLS yields no periods above the 99.9\% FAP line (see \reffig{fig:hip98599-periodogram}), the most prominent ones ($\geq99.0\%$) at 7.2~d, 25.5~d and 155.9~d. The residuals' correlogram also flattens, with $\rho_{\mathrm{RV, FW}}$=-0.09, $\rho_{\mathrm{RV, BIS}}$=0.17.

\begin{figure} 
    \includegraphics[width=\columnwidth]{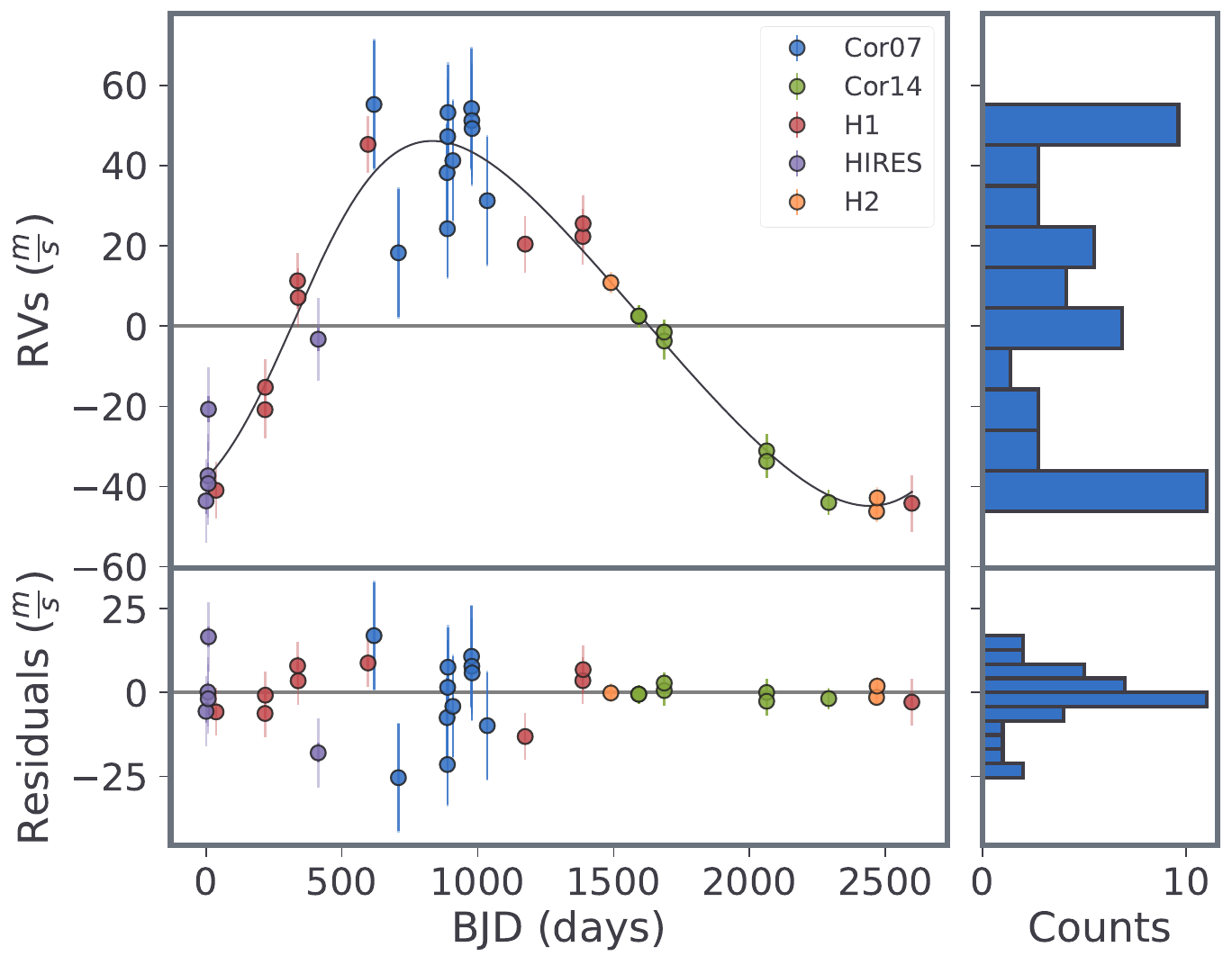}
    \caption{HIP\,98599 RVs phase-folded at $P_1=2656^{+40}_{-16}$~d as circles coloured per instrument with the \onek~model imposed as a black line. Bottom, residuals. Right of each plot, RVs histograms.}
    \label{fig:hip98599-keplerian}

    \includegraphics[width=\columnwidth]{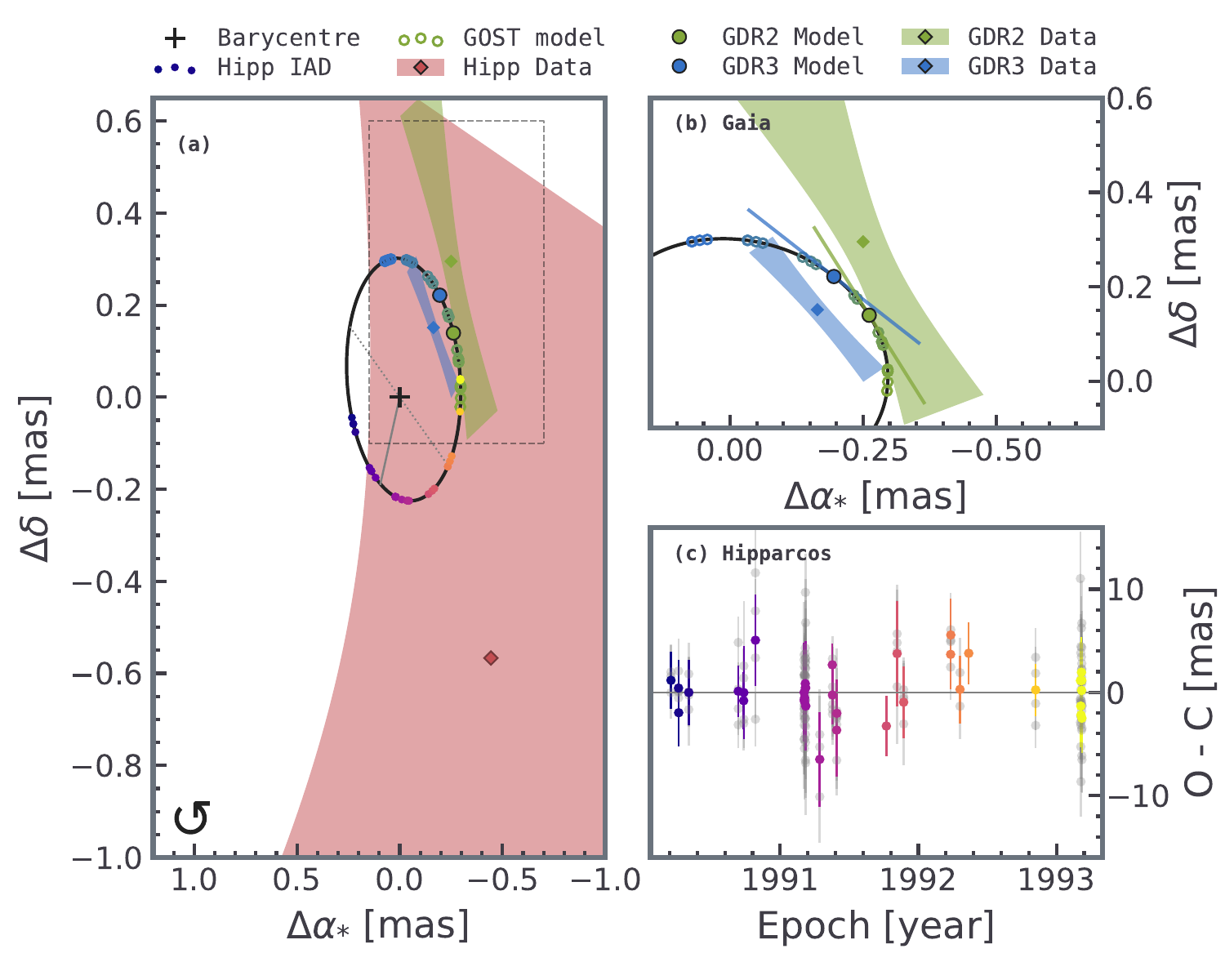}
    \caption{HIP\,98599 best-fit for the \onek~model. Displaying for panel (a) the best-fit astrometric orbit, (b) zoom-in for Gaia and GOST model, gray rectangle in panel (a), and (c) residuals for Hipparcos abscissa.}
    \label{fig:hip98599-plot_am}
\end{figure}

\section{Discussion} \label{sec:discussion}

The \emp~framework, with the addition of astrometric differencing modelling \citep{2019MNRAS.490.5002F, 2021MNRAS.507.2856F} introduced in Section \ref{sec:bayesian_framework_bayesian_approach}, proves to be a useful tool for identifying and characterising cold giants. In our analysis of this first batch from the extended CHEPS--totalling an RV baseline of $\sim$16~yrs--we validate our tool by refining orbital parameter estimates for HIP\,21850, and fully characterise 5 new exoplanets; a warm Jupiter (HIP\,10090c), and 4 Jupiter analogues (HIP\,8923b, HIP\,10090b, HIP\,98599b, and HIP\,39330b); as well as a warm Saturn candidate (HIP\,39330c), summarised in \reftab{tab:planets-summary}. In this section we will quantify how the inclusion of astrometry data boosts confidence in the planetary solutions via Bayesian Evidence, and helps with tighter uncertainties for orbital parameters. Then we discuss the multi-modality problem presented by HIP\,39330, and address how including small datasets affects our inference. Finally, we place the CHEPS planets of this study in context with the overall exoplanet population, and highlight future opportunities.

\subsection{Improved model evidence}\label{sec:discussion-model-evidence}

Adding astrometry resulted in consistent periods with those found by RVs alone, and also improved the Bayes' factor between models. Defining this increase as $\Delta\mathrm{BF} =\Delta$\lnzh$\mid_{\mathrm{RV+AM}}$-$\Delta$\lnzh$\mid_{\mathrm{RV}}$, we see a wide range of model comparison improvement; in HIP\,8923 we have $\Delta\mathrm{BF}=58.58$, whereas in HIP\,21850 we have $\Delta\mathrm{BF}=1.70$. To further analyse this increase, we briefly introduce two metrics for our data, a phase-coverage metric $\Phi_c(P) \in \left[0, 1\right[$ and a baseline-coverage metric $\beta_c \in \mathbb{R}^+$. For a given test period $P$, we map to phase each observation time $t_i$, with $i=(0,\dots, n)$ for $n$ observations, as
\begin{equation}
    \phi_i \equiv \left( \frac{t_i - t_{0}}{P}\right) \qquad (\bmod1) .
\end{equation}
Then, we sort the phases and compute the circular gap $g_i$:
\begin{equation}
    g_i = 
    \begin{cases}
        \phi_{i+1} - \phi_{i} & \text{for $i=1,\dots, n-1$}\\
        1-\phi_n + \phi_1 & \text{for $i=n$}
    \end{cases}\,.
\end{equation}

Each $g_i$ reveals the observational gap in phase. And with $\Phi_c\equiv \max(g_i)$, we have a metric for `phase-incompleteness', where $\Phi_c\rightarrow0$ translates to full period coverage (impossible for a finite sample), and $\Phi_c\rightarrow1$ to no phase coverage. We computed $\Phi_c$ for a period grid [0, 20\,000] for each data type and system, obtaining a periodogram-like figure (see \reffig{fig:phase-coverage}), where peaks mark the periods that have the least coverage. Additionally, with the baseline-coverage $B(t)$ (how many orbital cycles the baseline spans) we calculate a baseline-coverage improvement ratio $B_c$ as

\begin{equation}
B(t) = \frac{t_{\mathrm{max}} - t_{\mathrm{min}}}{P}\,, \qquad B_c=\frac{B(t_{\mathrm{RV+AM}})}{B(t_{\mathrm{RV}})}\,.
\end{equation}

We  use these metrics, as they capture the fraction of orbital phase spanned by observations ($\Phi_c$) and the extent to which astrometry extends the temporal baseline ($B_c$) in a intuitive way.

For HIP\,8923, we have $\Delta \mathrm{BF}=58.58$, a tremendous information jump by the inclusion of the astrometry data. This can be easily explained: for our period of interest ($P_1\approx5162$~d) $\Phi_c$ decreases considerably with the addition of astrometry data, from $\Phi_{c, \mathrm{RV}}=0.468$ to $\Phi_{c, \mathrm{RV+AM}}=0.194$. This can be seen by comparing the height ($\Phi_c$) of the blue line (RV) against the purple line (RV+AM), across the green vertical bar ($P_1\approx5162$~d) in \reffig{fig:phase-coverage}).
A summary of comparative values can be found in \reftab{tab:bayes-evidence}. Evidence uncertainties are all $\approx 0.2$, with $\Delta \mathrm{BF}$ uncertainty $\approx 0.3$. Exact evidence uncertainties can be found in Section \ref{sec:results}.

At the other extreme, HIP\,21850 has the smallest $\Delta \mathrm{BF}=1.70$, well explained by $\Phi_c$ barely increasing with astrometry $\Delta \Phi_c=0.016$ (in \reffig{fig:phase-coverage}, our period lies in a peak), and the lowest baseline-coverage ratio $B_{c}=1.687$.

HIP\,10090 has the second largest increase $\Delta \mathrm{BF}=16.30$. While not directly explained by coverage-metrics, in the RVs there is a secondary peak in the posteriors at 395~d, with $\Delta$likelihood $\sim$6 from the maximum, which is dispelled with astrometry. On the reverse side, HIP\,39330, while having good coverage-metrics, the addition of astrometry creates a multimodal solution, penalising the overall model likelihood, with the second lowest $\Delta \mathrm{BF}=3.04$. Looking at the phase-coverage provided by the astrometry in \reffig{fig:phase-coverage}, our long period resides right in the middle of a red peak $\Phi_{c, \mathrm{AM}}=0.723$, meaning astrometry does not provide good coverage at all. The short period (orange vertical line), while being almost completely covered $\Phi_{c, \mathrm{AM}}\sim0$, is almost imperceptible in the astrometry data (see the small wobble in \reffig{fig:hip39330-plot_am}). It is worth mentioning that the inclusion of astrometry, while mostly uninformative for the \twok\,model, does indeed help with the first signal. When comparing Bayes' factor increase between \zerok\,and \onek, the increase is of $\Delta \mathrm{BF}=9.30$.

\begin{table}[!ht]
\caption{\label{tab:bayes-evidence}Improvement of Bayes' factors (BF) and coverage metrics ($\Phi_c$ for phase and $B_c$ for baseline) with joint RV+astrometry over RV-only.}

\centering
\begin{tabular}{lllll}
\toprule
Planet    &Period (days) &$\Delta \mathrm{BF}$ & $\Delta \Phi_c$ & $B_{c}$ \\
\midrule
\midrule
HIP\,8923b   &$5160^{+150}_{-240}$ &58.58 &0.274 & 2.372\\
HIP\,10090b  &$2960^{+120}_{-100}$ &16.30 &0.062 & 2.068\\
HIP\,21850c  &$11320^{+490}_{-940}$&1.70  &0.016 & 1.414\\
HIP\,39330b  &$4650^{+210}_{-250}$ &3.04  &0.108 & 2.064\\
HIP\,98599b  &$2656^{+40}_{-16}  $ &13.62 &0.056 & 1.687\\
\bottomrule
\end{tabular}
\end{table}

\subsection{Improved planet parameter estimates}\label{sec:discussion-improved-planet-parameters}

Most of our systems present improved parameter estimates, for example, mass determination helps securing HIP\,10090c from an ambiguous Saturn \msini=$0.39^{+0.02}_{-0.03}~M_J$ to a bona-fide Jupiter-mass planet $M=0.85^{+0.03}_{-0.12}~M_J$. For HIP\,21850 we have tighter period constraints $P_1=2534^{+22}_{-11} \rightarrow 2539^{+1}_{-3}$~d, and for the second period $P_2=15900^{+2500}_{-4800} \rightarrow 11320^{+490}_{-940}$~d, reducing uncertainties by almost an order of magnitude, as well as near-coplanar orbits $\Delta I=7.2^{\circ}$ for this system.  For HIP\,98599 we reduce the period uncertainty range by a third $P_1=2580^{+80}_{-90} \rightarrow 2656^{+40}_{-16}$~d. 

For HIP\,39330, we are faced with an awkward situation, effectively leaving us with two astrometric solutions. This means our data is genuinely insufficient to identify a unique orbital configuration (see \reffig{fig:hip39330-cornerplot}), bifurcating the second orbital period into two distinct solutions, $P_{2,1}\approx255~d$ and $P_{2,2}\approx260~d$, and leaving us unable to resolve the inclination degeneracy. We fitted each mode separately with normal priors on the inclination $I_{1, 1}$\Normal{0.92}{0.3}, and $I_{1, 2}$\Normal{2.29}{0.3} and quantified their odds, which are virtually equal. While there is enough evidence to confidently claim a second signal, we can't confidently characterise it. Conservatively, we prefer adopting the simpler \onek~model, until there is further astrometric data to break this degeneracy. Parameter summaries for each mode can be found in \reftab{tab:HIP39330_params}.

\subsection{Sufficient RV data}

Our need to split RV datasets into subsets as small as $N=3$ (see \reftab{tab:n-obs-target}) raises the question: is the information obtained by so few points worth the additional offset parameter? We examine in detail HIP\,10090, composed of a grand total of 57 RVs--12 COR07, 3 COR14T, 6 COR14, 32 H, and 4 H15 (datasets defined in \refsec{sec:observations}). Out of these datasets, COR07, COR14T, and H15 do not provide additional baseline. Furthermore, their RV phase-coverage is shared with other datasets.  We compare the full RV phase-incompleteness against the RVs without each one of these sets (which we denote with the sub-indices $\mathrm{-C07}, \mathrm{-C14T}, \mathrm{-H15}$), obtaining for the long period $P_1=2960^{+120}_{-100}$ the values $\Phi_{c, \mathrm{RV}}=0.163$, $\Phi_{c, \mathrm{-C07}}=0.213$, $\Phi_{c, \mathrm{-C14T}}=0.163$, and $\Phi_{c, \mathrm{-H15}}=0.163$, and for the short period $\Phi_c$=0.136, 0.158, 0.136, and 0.141, for all combined datasets, and -COR07, -COR14T, and -H15, respectively. With the exception of COR07 for the long period, none of these sets have unique phase-coverage, reflected in the phase-coverage metric staying at 0.163 when removing datasets.
Several interesting points rise up when examining the evidences (see \reftab{tab:hip10090-incomplete-rvs}). Out of these subsets, COR07 drives the most evidence, presenting the highest evidence gap with respect to the full RVs $\sim$45, for both \onek~and \twok. Expectedly, the solution loosens up for the long period $P_1=2920_{-120}^{+180}$~d, while remaining with uncertainties under unity for $P_2=321.38_{-0.23}^{+0.69}$~d. Nonetheless, this information does not contribute to the model comparison, with similar $\Delta$\lnzh~to the full RVs $\Delta$\lnzh=$27.22\approx27.65$.
A similar phenomenon occurs for H15, but driven by the high precision of the data instead of its quantity, where the model comparison stays similar ($\Delta$\lnzh=27.22) with the long period uncertainty increasing when removing H15 $P_1=3250_{-230}^{+130}$~d, and $P_2=320.71_{-0.04}^{+1.04}$~d staying similarly constrained. 
On the other hand, and perhaps unexpectedly, COR14T's three measurements, while contributing the least to the overall evidence of \twok~with an almost marginal difference of $1.43$ against the full-RV, have a crucial role in model comparison, driving down the $\Delta$\lnzh to 11.88 when removed (from 27.22). These three modest measurements have the peculiarity of overlapping with GOST data, driving both the inclination and the astrometric offset posteriors to a tighter solution.

\begin{table}[!ht]
\caption{\label{tab:hip10090-incomplete-rvs}Bayes' factors and coverage metrics summary.}

\centering
\begin{tabular}{lllll}
\toprule
Planet    &$\Phi_{c, \mathrm{RV}}$ & \lnzh$\mid_{1\mathcal{K}}$& \lnzh$\mid_{2\mathcal{K}}$& $\Delta$\lnzh \\
\midrule
\midrule
FULL      &0.163 & -562.79&-535.57 &27.22  \\
-COR07    &0.213 & -517.67&-490.01 &27.65  \\
-COR14T   &0.163 & -546.02&-534.14 &11.88  \\
-H15       &0.163 & -552.12&-525.43 &26.69  \\
-COR14T,H15&0.163 & -535.35&-524.00 &11.36  \\
\bottomrule
\end{tabular}
\end{table}

\subsection{Cold Jupiters and next-steps}

Our CHEPS analysis places four systems firmly in the Jupiter analogue regime (masses of 0.3-13$M_J$ on orbits of 3-7~AU) and provides dynamical characterisation for each with modest eccentricities $e\lesssim0.19$ (see \reftab{tab:planets-summary}). Two of the planets, HIP\,8923b and HIP\,39330b, reside very close to Jupiter's orbital scale ($\sim$5~AU), directly probing Solar-System-like architectures. Furthermore, for HIP\,39330c, although we could not precisely constrain the inclination with our current data, it resides at $\sim$1~AU, making it a Jupiter-Earth-like system.

\begin{figure}[!ht]
    \includegraphics[width=1.1\columnwidth]{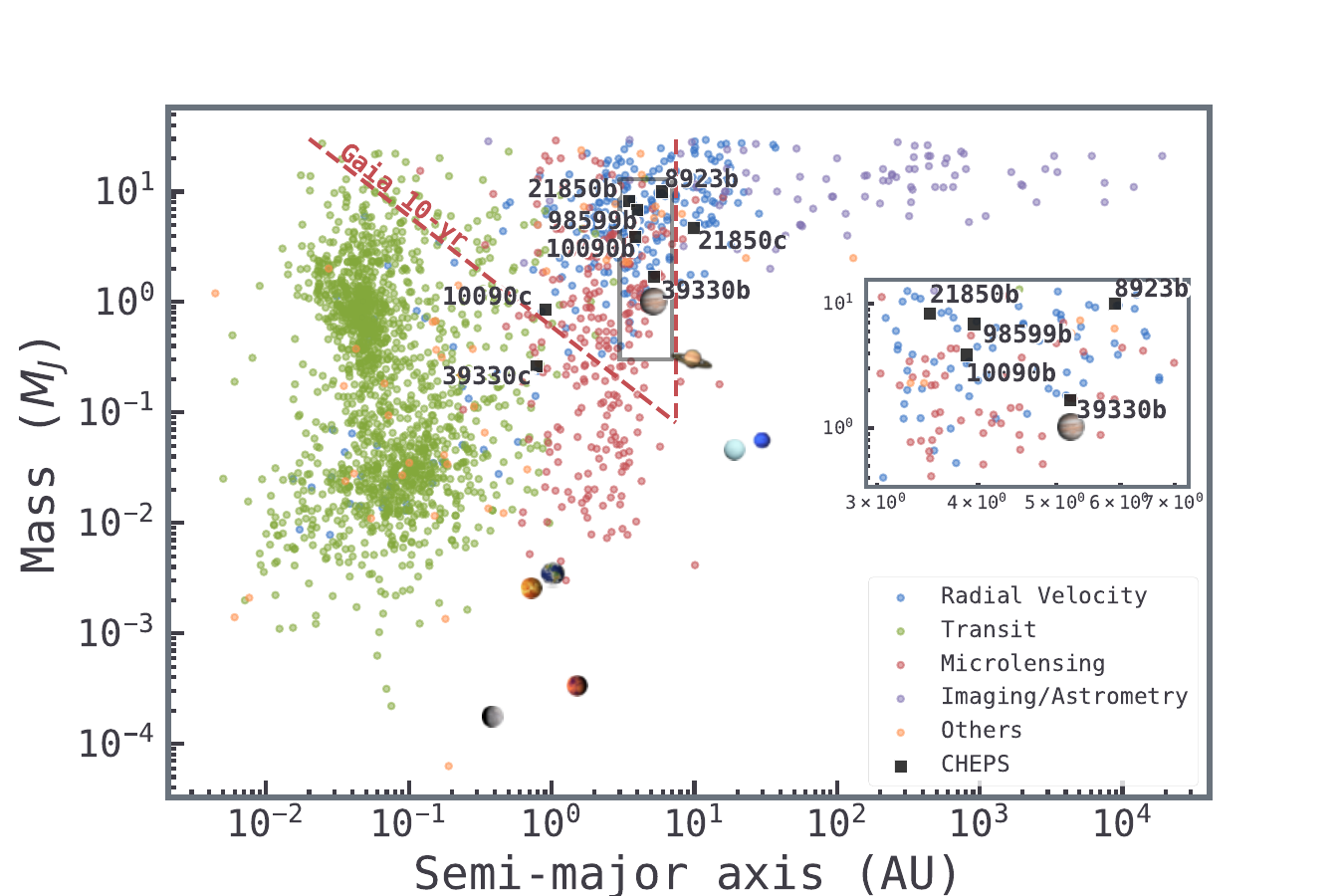}
    \caption{True mass vs semi-major axis of the confirmed exoplanet population from the NASA Exoplanet Archive \citep{2025PSJ.....6..186C}, coloured by discovery method. The approximate sensitivity curve for GDR4 is shown in dashed red, assuming a Sun-like star at 20~pc and a $3\sigma$ detection criterion. CHEPS targets in this study are placed as solid black squares. Solar-System planets are placed as icons.}
    \label{fig:planet-pop}
\end{figure}

Of the more than 6\,000 confirmed exoplanets to date, only about a third have true mass estimates \citep[2\,223 according to ][]{2025PSJ.....6..186C}. Most of these come from transits (1\,567), so they predominantly sample short-to-intermediate orbital periods. \reffig{fig:planet-pop} situates these planets within the broader exoplanet census in the mass-distance plane. The CHEPS Jupiter analogues occupy the transitional swath where RV sensitivity declines with increasing $a$ and direct imaging has focused beyond several AU. By combining decades of precise RV monitoring with Hipparcos-to-Gaia astrometry, we are now filling this demographic gap with dynamical masses instead of minimum masses. This expands the well-vetted sample of true Jupiter analogues, improving constraints for population-level inferences and forward models of direct-imaging yields. Furthermore, the other two planets (HIP\,10090c and HIP\,39330c) populate a far more devoid area, around $\sim$1~AU.

Beyond the characterisation of individual systems, joint RV+astrometry inference is rapidly becoming central to catalogue-level reliability. In practice, Gaia will turn many long-period RV detections into directly testable hypotheses against Gaia astrometry, either reinforcing the planetary interpretation or exposing tension that points to additional companions, alternative solutions, or underestimated systematics. This shift will also deliver a step-change in mass completeness, moving a large fraction of long-period companions from minimum to true masses and thereby enabling cleaner demographics of cold giants and a more robust separation of planets from brown dwarves and low-mass stellar companions. Our Hipparcos--Gaia differencing and joint-modelling approach already mirrors, in compressed form, the type of information Gaia DR4/DR5 will provide at scale through intermediate astrometry and non-single-star solutions, so the same modelling philosophy and vetting logic will carry over naturally. At the same time, Gaia-based candidate lists will inevitably include astrophysical contaminants (e.g., unresolved binaries), strengthening the case for joint, physically consistent modelling rather than naive catalogue ingestion.

Looking ahead, these systems are ripe for targeted follow-up. Extended RV baselines will continue to refine long-period phase-coverage, while upcoming astrometric surveys and data releases, like the Rubin Observatory Data Release 1 on May 2026 \citep[LSST,\,][]{2023PASP..135j5002H}, Gaia's DR4 on December 2026 \citep{2024eas..conf..208B}, the Roman Space Telescope in late 2026-2027 \citep{2025arXiv250814412H}, or CHES \citep{2024ChJSS..44..193J}, will sharpen accelerations and proper-motion anomalies, or even determine orbits on their own with Intermediate Data. The same RV+astrometry framework that proved decisive here is directly aligned with near-term and mid-term roadmaps for cold giant demographics and precursor target vetting for future missions. Therefore we encourage the early adoption of available astrometric data in RV characterisation, providing full orbital parameters, as well as enhanced confidence in the modelled solutions, even when RVs alone might appear sufficient. In this context, CHEPS' metal-rich, quiet FGK targets remain an efficient hunting ground for Solar-System analogues that can anchor comparative planet studies across techniques. Building on CHEPS' 16-year survey of $\sim$240 stars, already yielding cold giants, we are extending the observational baseline with continued monitoring. This will expand the sample of Jupiter analogues with true masses and full orbits, enabling a robust population-level census of Jupiter analogues.

\section{Conclusions}  \label{sec:conclusions}

This study extended the Chile-Hertfordshire ExoPlanet Survey (CHEPS) by combining long baseline RV data with absolute astrometry from the Hipparcos and Gaia missions to search for cold, wide-orbiting gas giants. Using the renewed \emp~framework enhanced with astrometric differencing, we have analysed five metal-rich FGK hosts, and performed Bayesian model comparisons to identify planetary signals and quantify the improvement brought by astrometry.

The analysis confirmed the Jupiter-mass companions for HIP\,21850, refined their orbital parameters with CHEPS data--extending the RV-baseline from $\sim$16.4~yr to $\sim$24.6~yr--and discovered four new Jupiter analogues: HIP\,8923b, HIP\,10090b, HIP\,39330b, and HIP\,98599b. We also identified a new warm Jupiter, HIP\,10090c. True masses were derived thanks to the astrometric inclination constraints, yielding estimates from $\sim$1 to $\sim$10~$M_J$ on orbits between $\sim$1 and $\sim$7~AU for new planets (see \reftab{tab:planets-summary}). In one system (HIP\,39330), the astrometric data lead to a multimodal solution for the second planet; adopting a simpler one planet model is prudent until additional data becomes available.

Incorporating astrometric data increased Bayes' factors ranging from $\sim$1.7 to $\sim$58.6, depending on phase coverage, baseline extension, and posterior multimodality (see \reftab{tab:bayes-evidence}). Custom phase-coverage metrics we introduced show that astrometry significantly reduces gaps in the orbital phase for cold giants and more than doubles the temporal baseline for some systems. This leads to substantial reductions in period and mass uncertainties, with up to an order-of-magnitude improvement for certain parameters. Even sparse astrometric measurements, when combined with RVs, can break the $\sin{I}$ degeneracy and convert minimum masses into true masses.

The sufficiency test on HIP\,10090 illustrates that not all RV subsets contribute equally. While some high-precision but phase-redundant datasets have marginal impact, others with only a few observations can substantially influence the Bayes' factor if they sample unique orbital phases. This underscores the need for strategically timed RV observations to complement astrometric data. The inability to constrain the inclination of HIP\,39330c also highlights the current limitations of GDR2/GDR3 astrometry for certain orbital periods; Gaia DR4, Rubin DR1 and future missions will provide decisive improvements.

The CHEPS discoveries fill a demographic gap in the exoplanet census, occupying the transition region between RV-dominated detections and direct imaging (see \reffig{fig:planet-pop}). Establishing a sample of Jupiter analogues with dynamically measured masses will improve population statistics and refine predictions for direct imaging yields. Continued RV monitoring and forthcoming astrometric releases will tighten orbital constraints and possibly reveal additional companions. The RV+astrometry framework developed in this work is directly relevant to the nearby targets accessible to ELT coronagraphy and Roman Space Telescope; astrometry provides true masses, orbital geometry, and improved ephemerides that enable scheduling of direct detections and atmospheric characterisation of these cold worlds that reside at the upper boundary of the planetary mass space.


By combining long-term RV monitoring with absolute astrometry, this work demonstrates a powerful avenue for uncovering and characterising cold giant planets on Solar-System-like orbits. The framework presented here positions CHEPS and similar surveys to exploit future data releases, paving the way toward a more complete understanding of planetary architectures around nearby stars.

\begin{acknowledgements}
    All benchmarks were performed on a computer with an \emph{AMD Ryzen Threadripper 3990X 64-Core Processor} with 128Gb of DDR4 3200MHz RAM, limiting the use to 24 physical cores and 24 threads. Typesetting was carried out in \texttt{Overleaf} \citep{Overleaf2025}.
    PAPR and JSJ gratefully acknowledge support by FONDECYT grant 1240738 and from the ANID BASAL project FB210003. We also greatly appreciate the support from the CASSACA China-Chile Joint Research Fund through grant CCJRF2205. Part of this work was supported by the United States Fulbright Fellowship in partnership with the Fulbright Chile Commission, and ANID/Fondo ALMA 2024/No.31240039 ``On the origin of Warm-Giant Planets". We thank the anonymous referee for insightful suggestions that enhanced the quality of this manuscript.
\end{acknowledgements}

\bibliographystyle{aa}
\bibliography{cheps_am.bib}

\begin{appendix}

\onecolumn
\section{Periodograms}\label{sec:appendix_periodograms}

\begin{figure}[!ht]
    \begin{minipage}[b]{0.49\columnwidth}
        \includegraphics[width=\columnwidth]{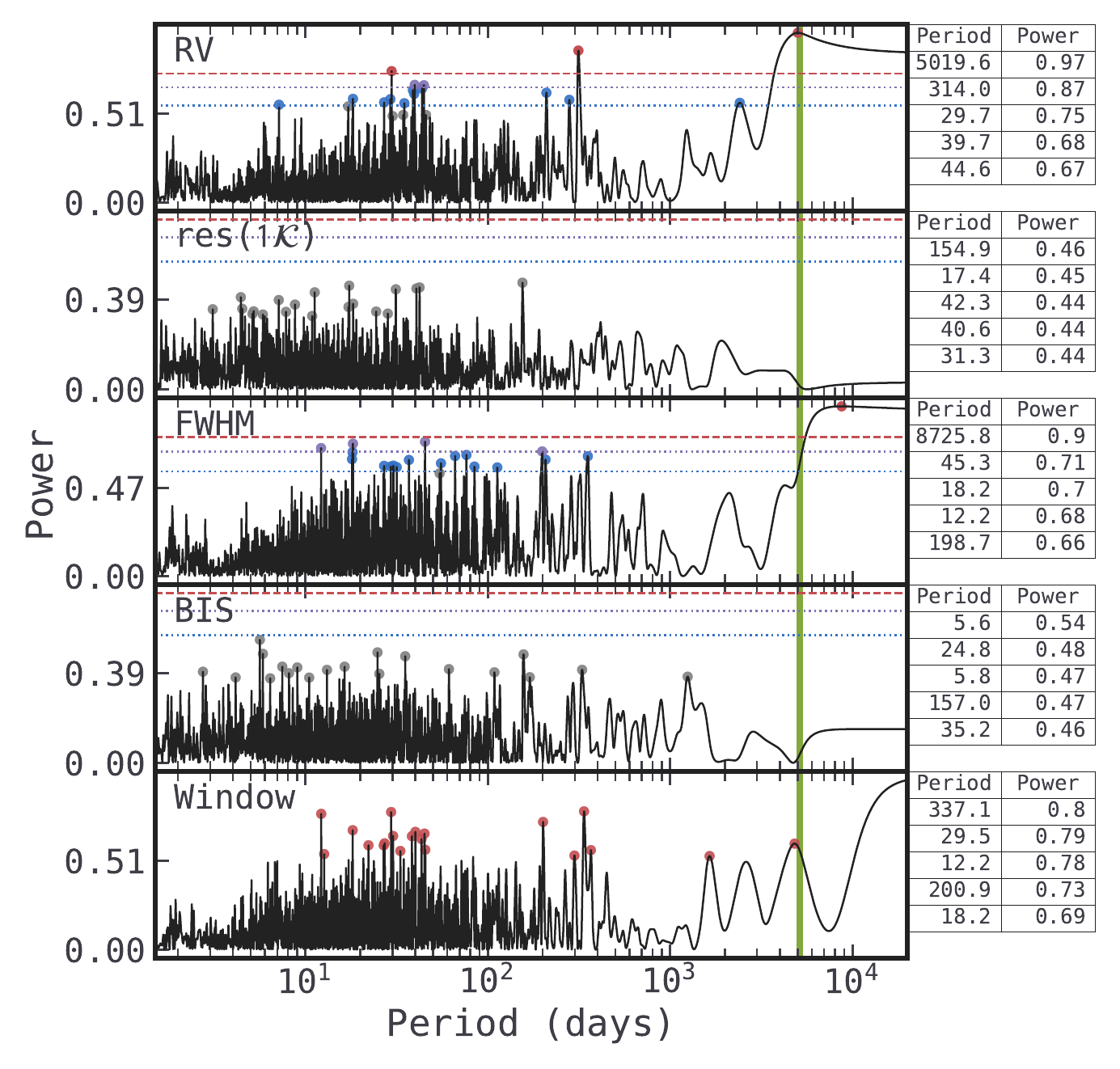}
        \caption{\label{fig:hip8923-periodogram}HIP\,8923 periodogram. Descending, RVs, model residuals, FWHM, BIS, and window function. FAP lines for 0.1, 1, and 10\%, in dashed red, dotted purple, and dotted blue, respectively. Circle markers show the periods with greatest power, coloured by FAP region. Green vertical region shows $P_1=5160^{+150}_{-240}$~d. On the right side there is a table summary of the five highest powers.}
    \end{minipage}\hfill
    \begin{minipage}[b]{0.49\columnwidth}
        \includegraphics[width=\columnwidth]{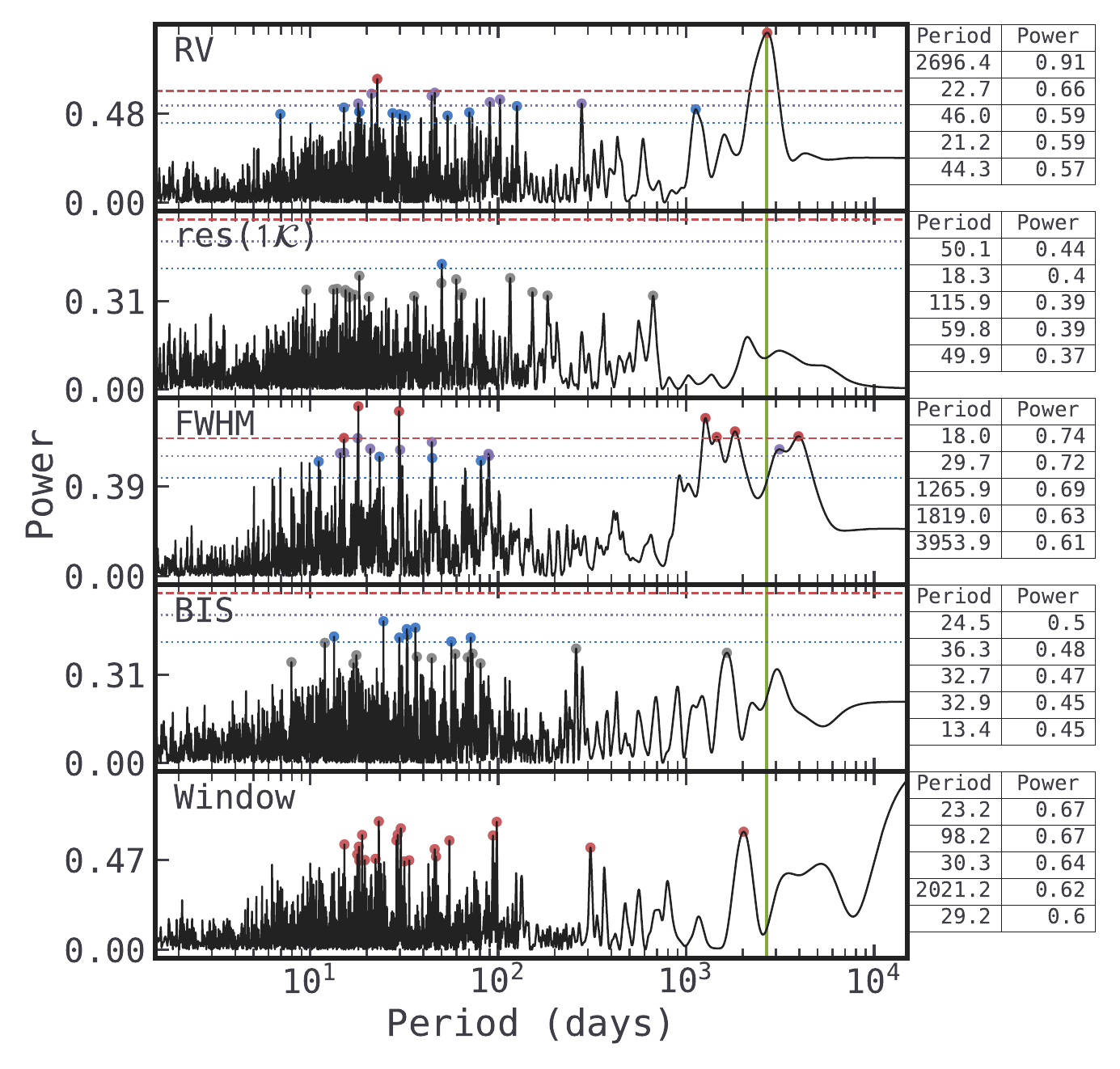}
        \caption{\label{fig:hip98599-periodogram}
        HIP\,98599 periodogram. Descending, RVs, model residuals, FWHM, BIS, and window function. FAP lines for 0.1, 1, and 10\%, in dashed red, dotted purple, and dotted blue, respectively. Circle markers show the periods with greatest power, coloured by FAP region. Green vertical region shows $P_1=2656^{+40}_{-16}$~d. On the right side there is a table summary of the five highest powers.}
    \end{minipage}
\end{figure}

\begin{figure}[!ht]
    \begin{minipage}[b]{0.49\columnwidth}
        \includegraphics[width=0.93\columnwidth]{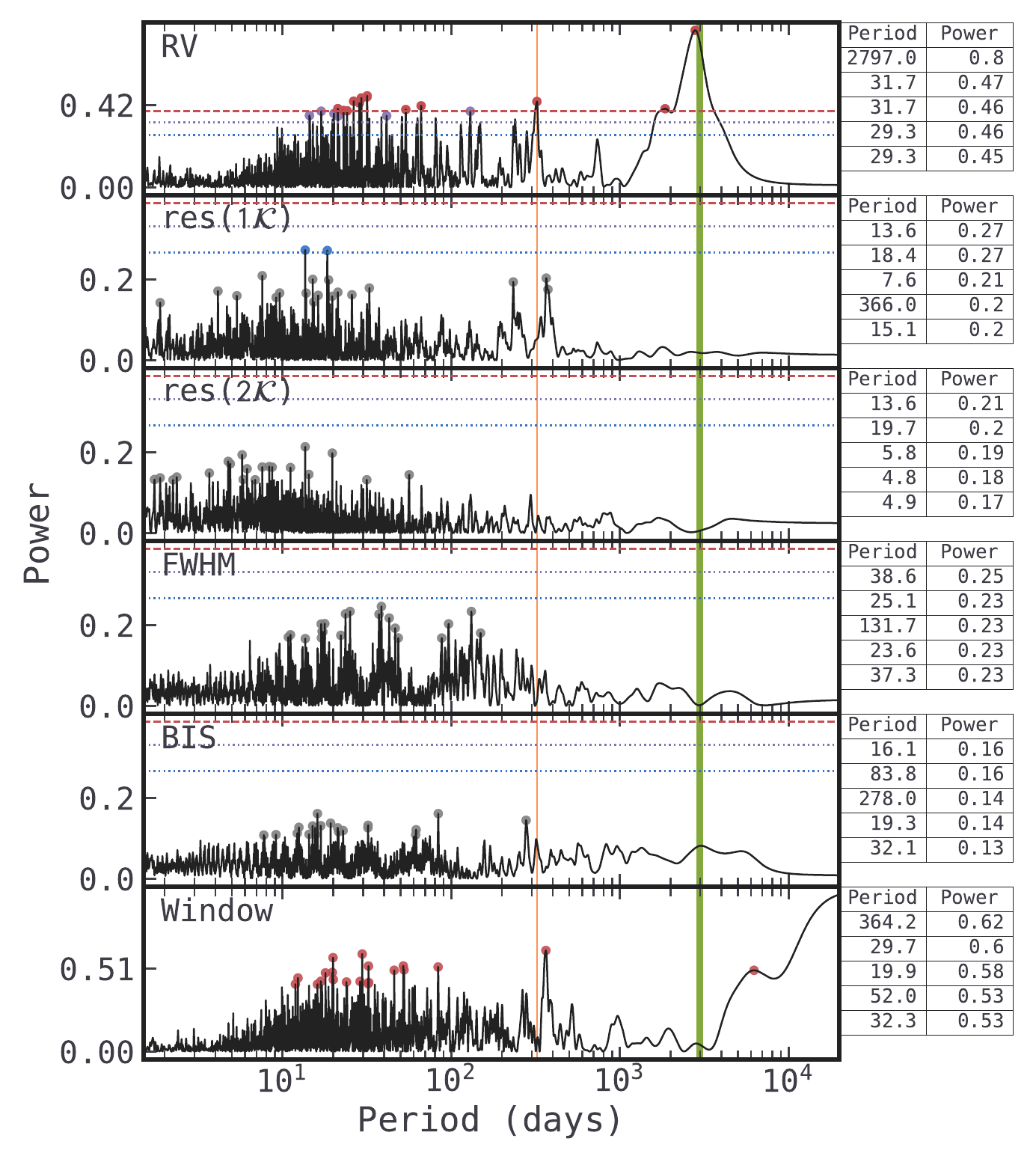}
        \caption{\label{fig:hip10090-periodogram}HIP\,10090 periodogram. Descending, RVs, model residuals, FWHM, BIS, and window function. FAP lines for 0.1, 1, and 10\%, in dashed red, dotted purple, and dotted blue, respectively. Circle markers show the periods with greatest power, coloured by FAP region. Green and orange vertical regions show $P_1 = 2960^{+120}_{-100}$~d and $P_2=321.82^{+0.32}_{-0.56}$~d, respectively. On the right side, a table summary with the highest powers.}
    \end{minipage}\hfill
    \begin{minipage}[b]{0.49\columnwidth}
        \includegraphics[width=0.93\columnwidth]{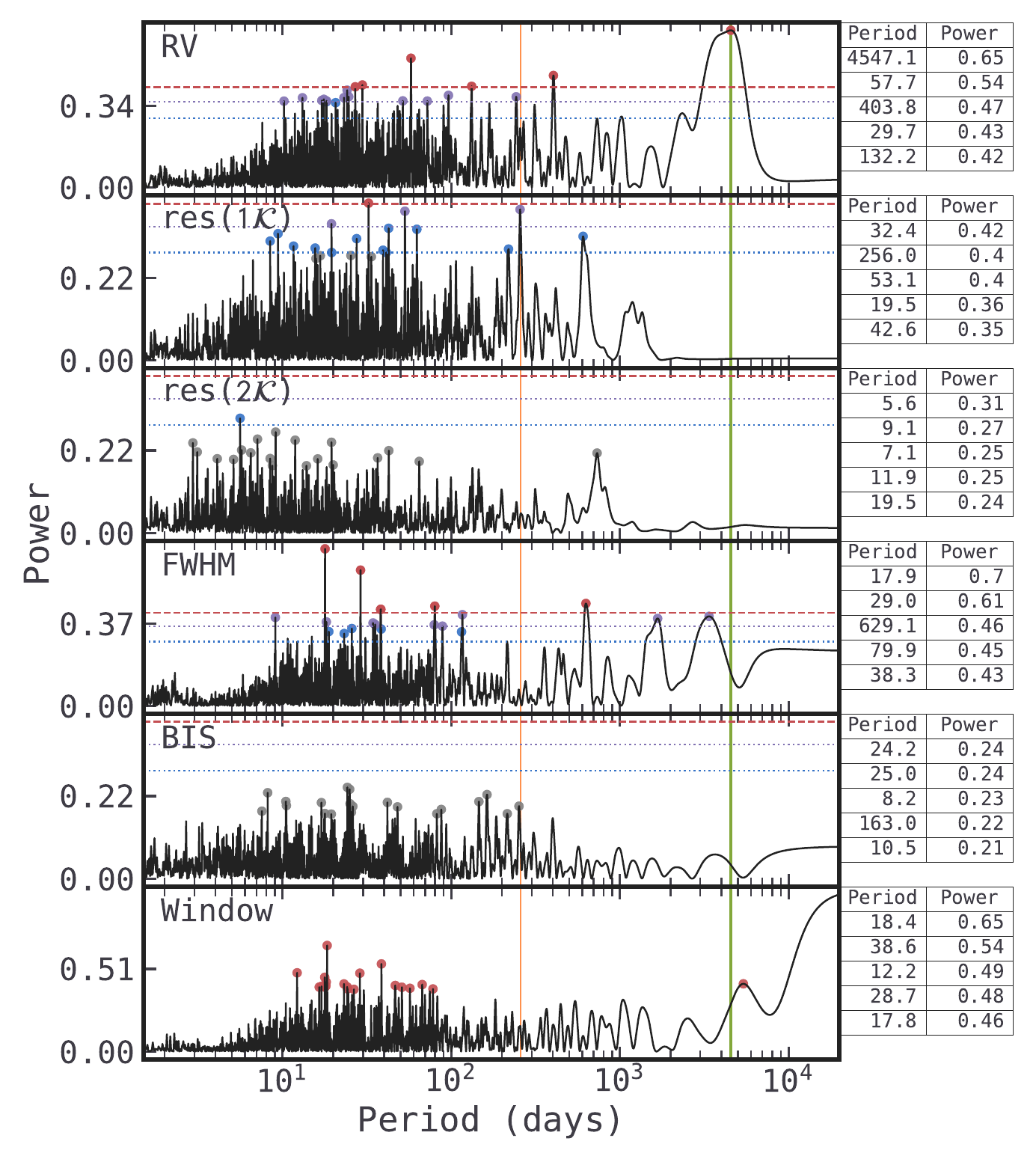}
        \caption{\label{fig:hip39330-periodogram}HIP\,39330 periodogram. Descending, RVs, model residuals, FWHM, BIS, and window function. FAP lines for 0.1, 1, and 10\%, in dashed red, dotted purple, and dotted blue, respectively. Circle markers show the periods with greatest power, coloured by FAP region. Green and orange vertical regions show $P_1 = 4596.9^{+1.7}_{-154.0}$~d and $P_2=257.1^{+1.6}_{-1.4}$~d, respectively. On the right side, a table summary with the highest powers.}
    \end{minipage}
    
\end{figure}

\onecolumn
\section{Correlograms}\label{sec:appendix_correlograms}

\begin{figure}[!ht]
  \centering
  \begin{minipage}[b]{0.49\columnwidth}
    \includegraphics[width=\columnwidth]{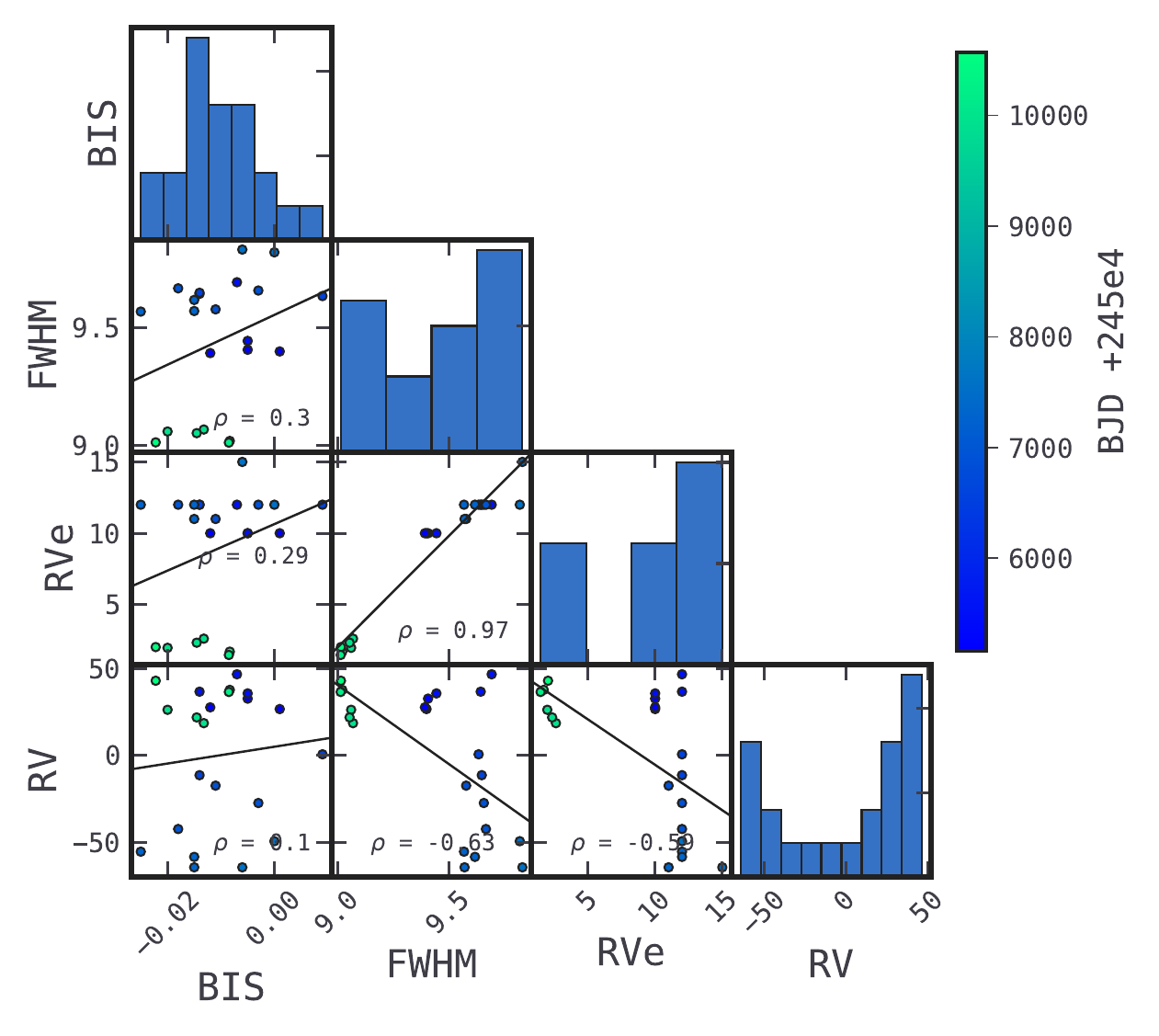}
    \caption{\label{fig:hip8923-correlogram}HIP\,8923 correlogram. Each panel displays the pairwise relation between RVs, full-width half-maximum (FWHM), CCF bisector inverse slope (BIS), and formal RV uncertainty (RVe).  Pearson correlation coefficient is denoted by $\rho$, with the linear trend as a black line.}
  \end{minipage}
  \hfill
  \begin{minipage}[b]{0.49\columnwidth}
    \includegraphics[width=\columnwidth]{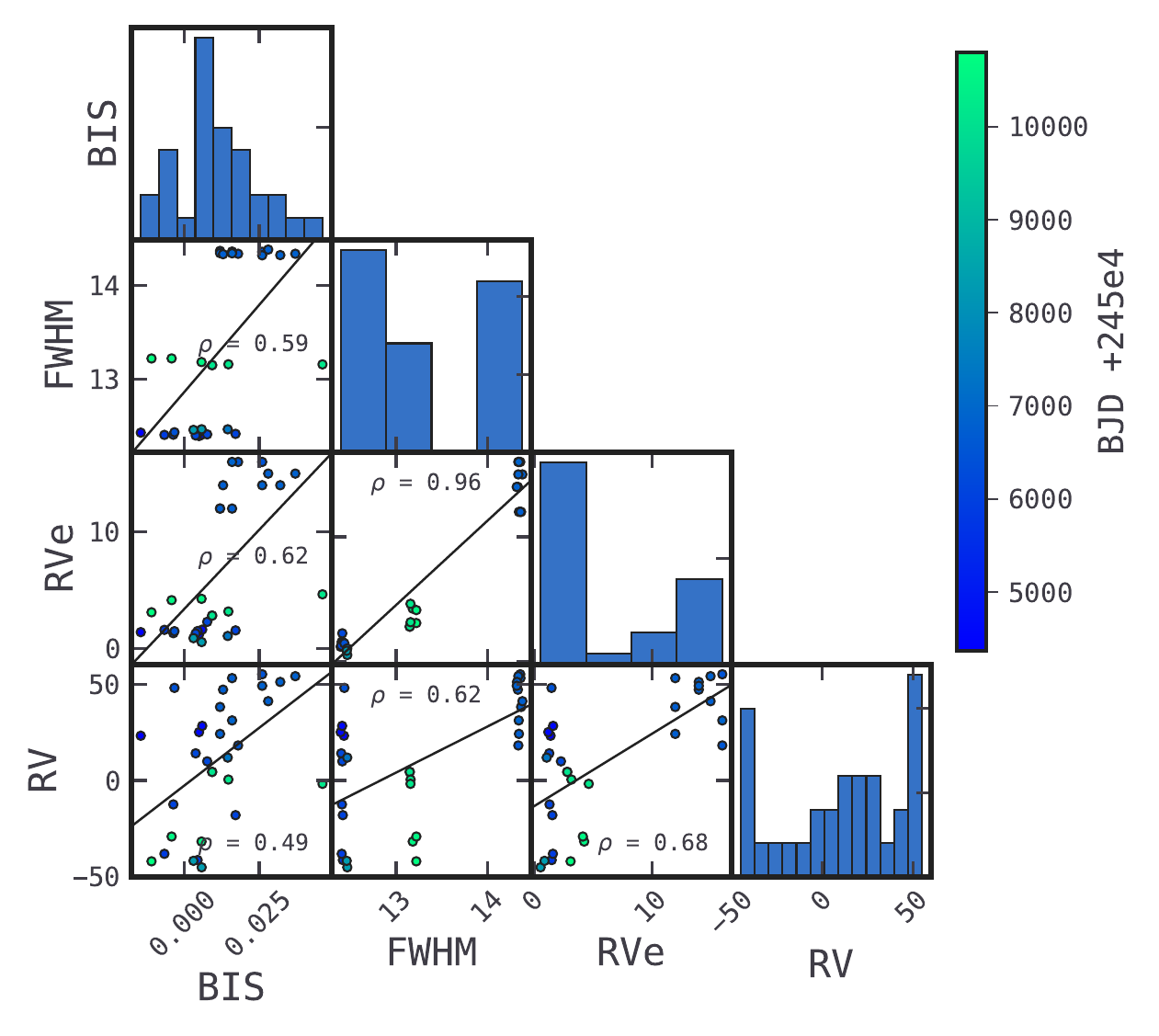}
    \caption{\label{fig:hip98599-correlogram}HIP\,98599 correlogram. Each panel displays the pairwise relation between RVs, full-width half-maximum (FWHM), CCF bisector inverse slope (BIS), and formal RV uncertainty (RVe).  Pearson correlation coefficient is denoted by $\rho$, with the linear trend as a black line.}
  \end{minipage}
\end{figure}

\begin{figure}[!ht]
  \centering
  \begin{minipage}[b]{0.49\columnwidth}
    \includegraphics[width=\columnwidth]{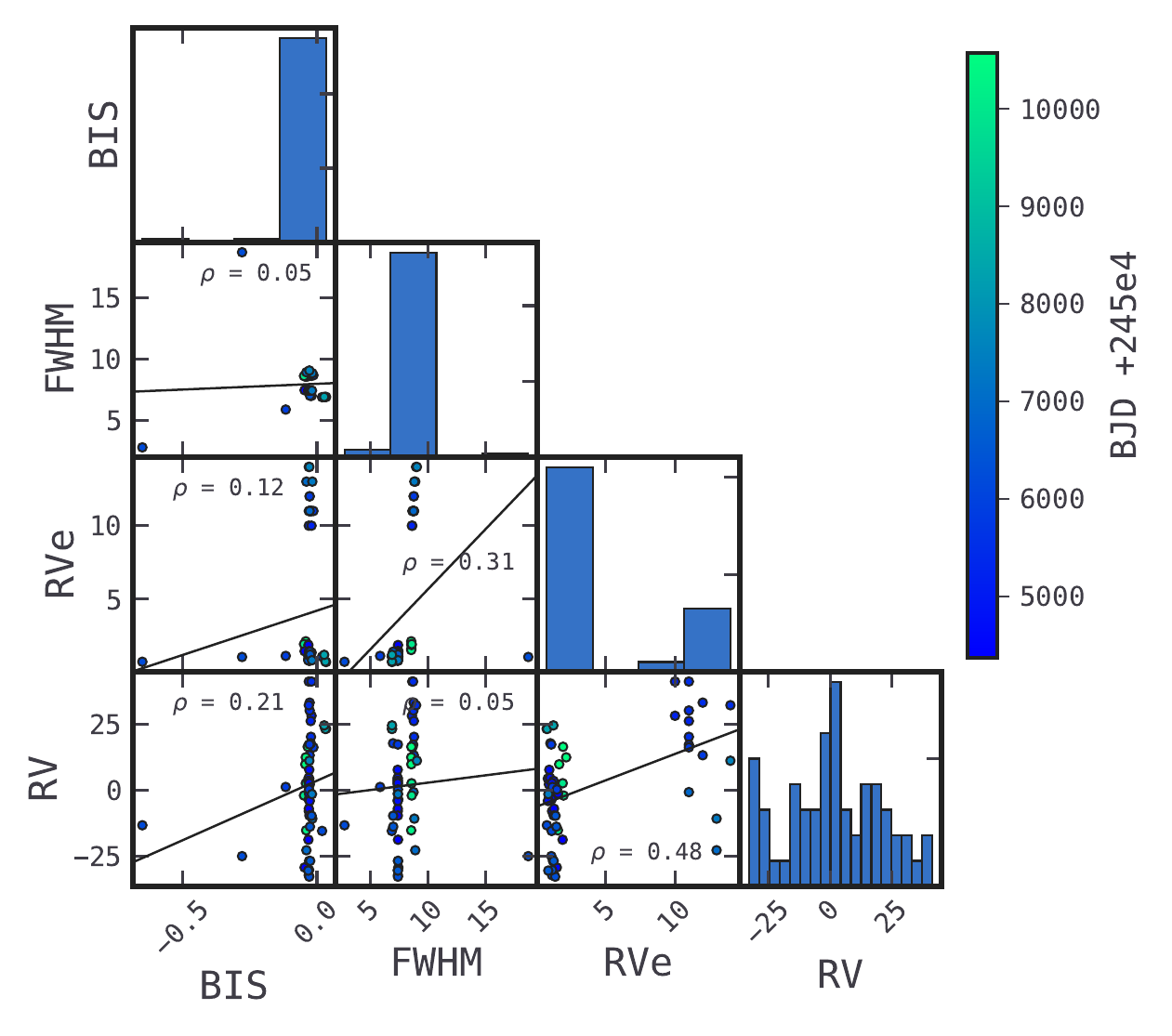}
    \caption{\label{fig:hip10090-correlogram}HIP\,10090 correlogram. Each panel displays the pairwise relation between RVs, full-width half-maximum (FWHM), CCF bisector inverse slope (BIS), and formal RV uncertainty (RVe).  Pearson correlation coefficient is denoted by $\rho$, with the linear trend as a black line.}
  \end{minipage}
  \hfill
  \begin{minipage}[b]{0.49\columnwidth}
    \includegraphics[width=\columnwidth]{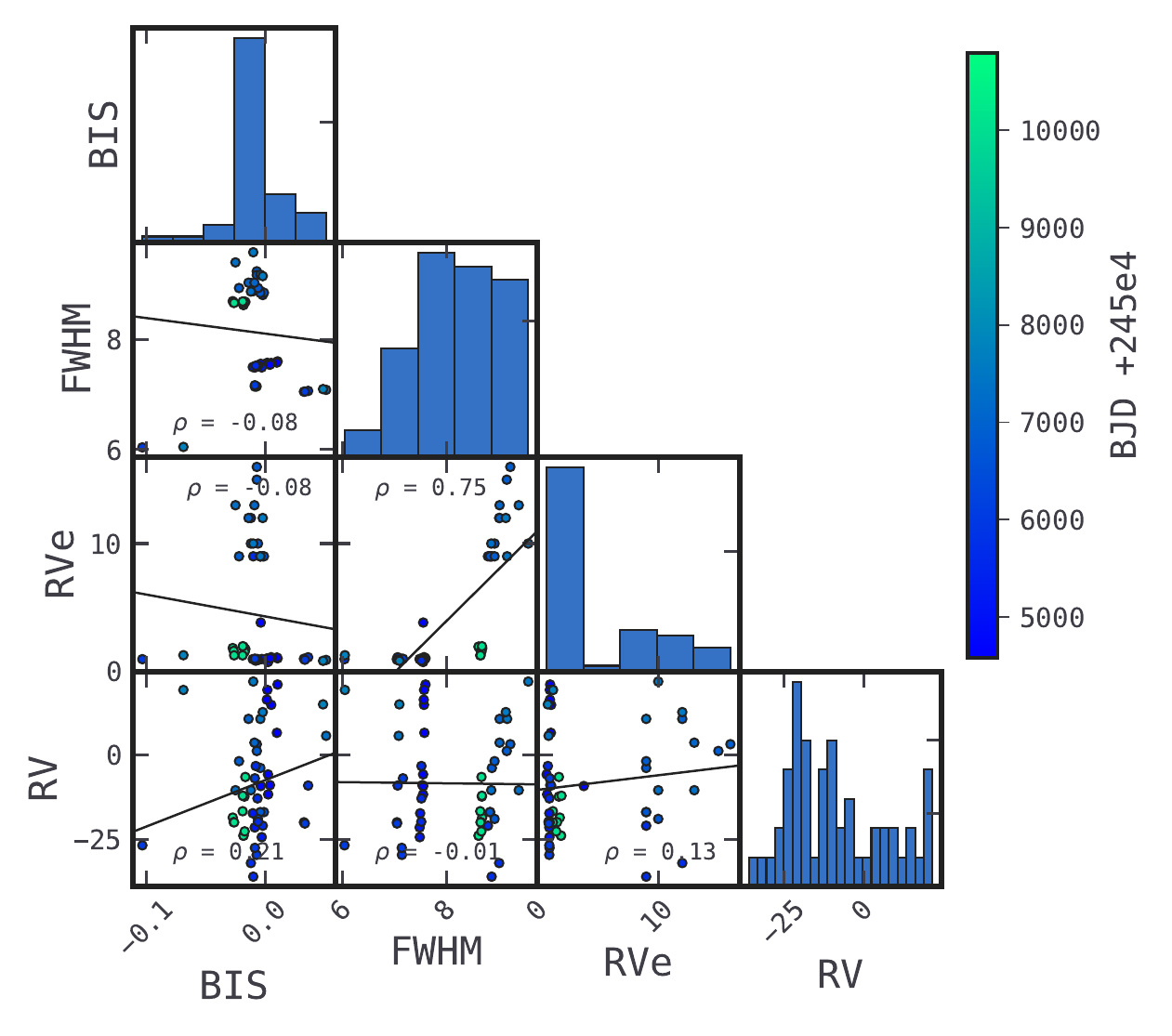}
    \caption{\label{fig:hip39330-correlogram}HIP\,39330 correlogram. Each panel displays the pairwise relation between RVs, full-width half-maximum (FWHM), CCF bisector inverse slope (BIS), and formal RV uncertainty (RVe).  Pearson correlation coefficient is denoted by $\rho$, with the linear trend as a black line.}
  \end{minipage}
\end{figure}


\onecolumn
\newpage
\section{More plots}\label{sec:appendix_rv_model}


\begin{figure}[ht!]
        \includegraphics[width=\columnwidth]{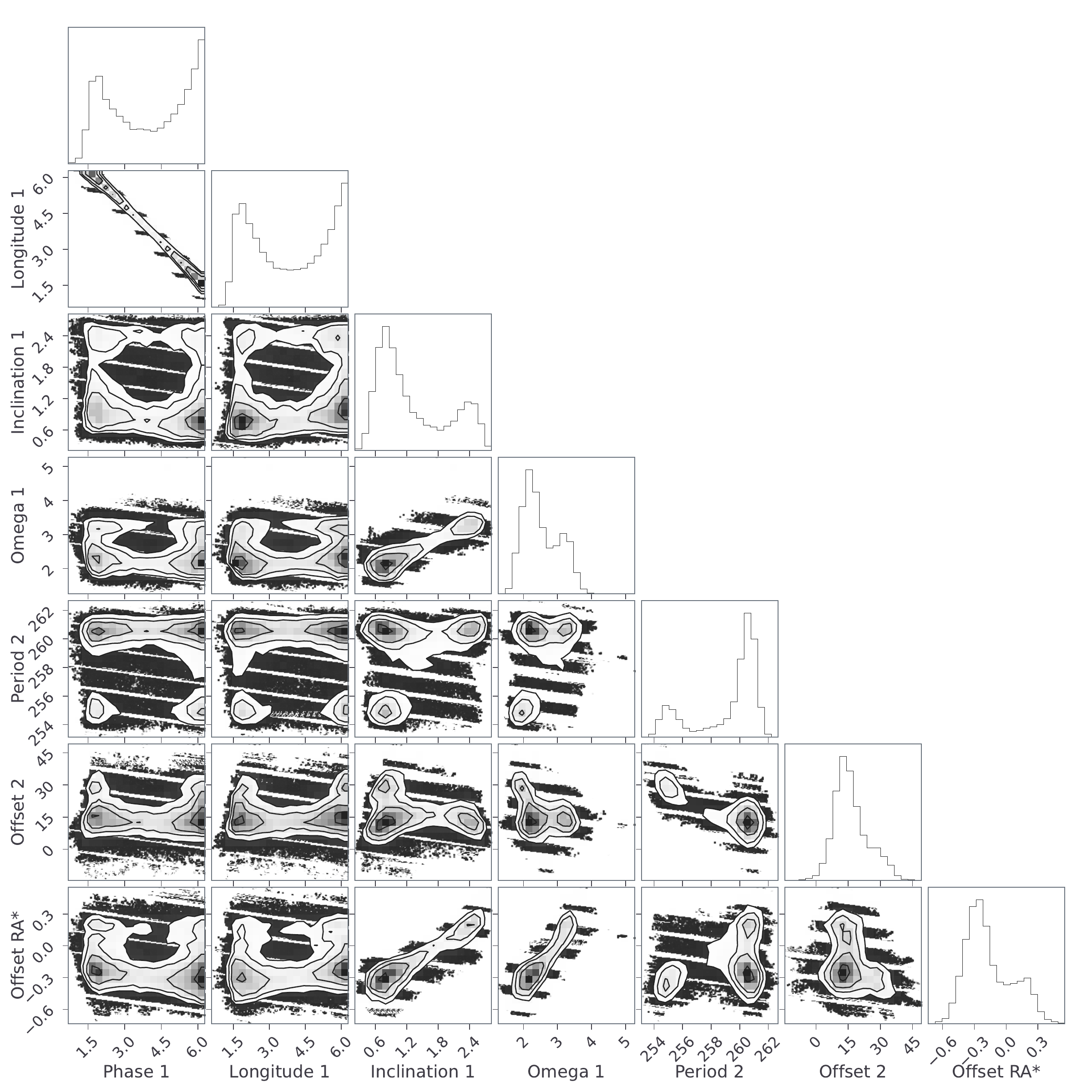}
        \caption{\label{fig:hip39330-cornerplot}HIP 39330 corner plot of selected parameters, displaying a bimodal exploration.}
\end{figure}

\newpage

\begin{figure}
\centering
\includegraphics[width=0.75\columnwidth]{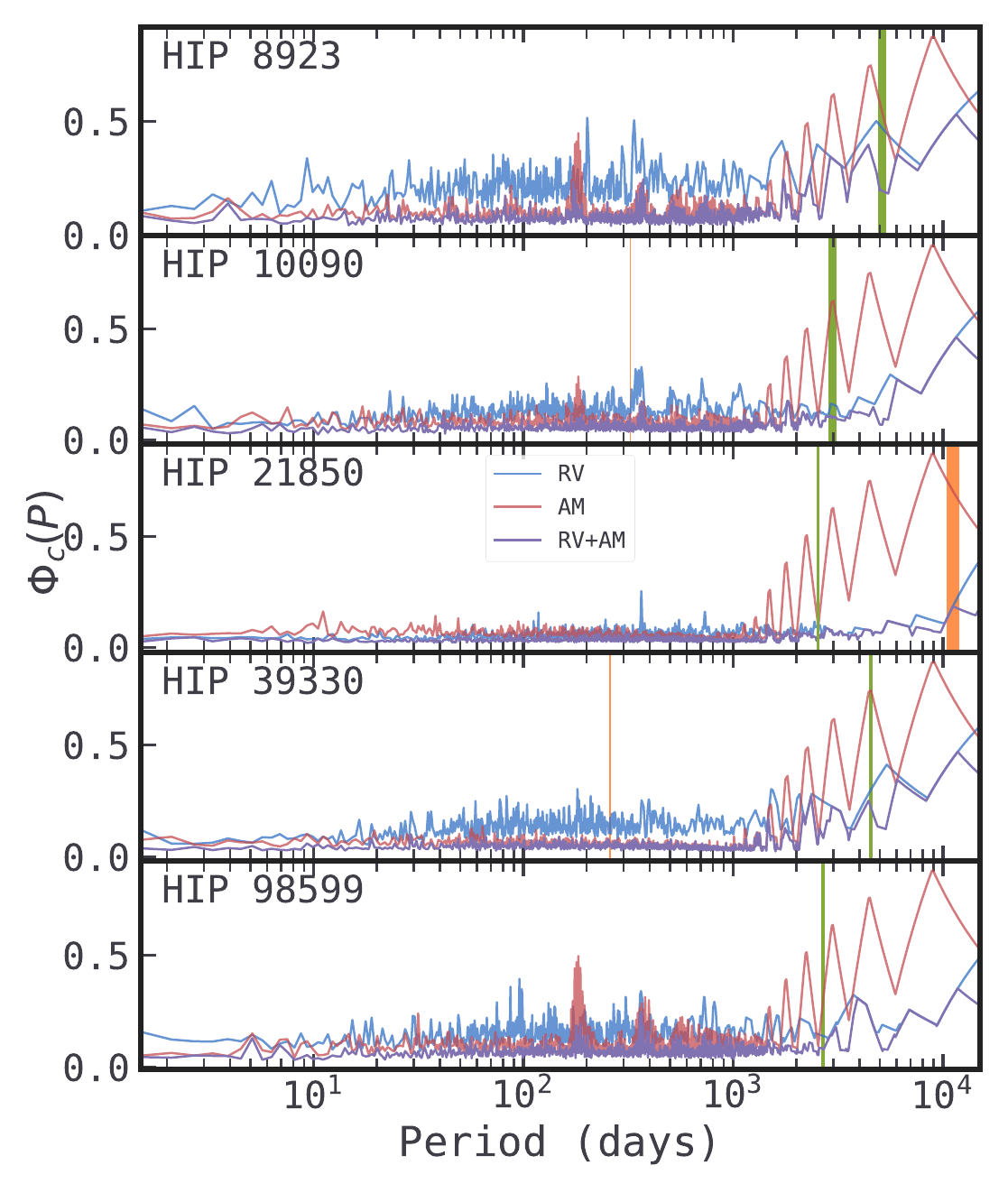}
    \caption{\label{fig:phase-coverage}Phase-coverage per system per data type. The y-axis shows phase-incompleteness $\Phi_c(P)$ as a function of period, with 0 meaning no gaps, increasing up to 1 with gap-size. Blue solid line corresponds to the RV data phase-coverage, red line to astrometry data, and purple to both combined. Green and orange vertical bars display the best-fit orbital periods.}
\end{figure}

\begin{table}
\centering
\caption{\label{tab:n-obs-target}Measurements per instrument subset (counts).}
\small
\setlength{\tabcolsep}{3pt}
\newcommand{\naa}{\textemdash}
\begin{tabular}{lcccccc}
\toprule
Target & COR (07/14T/14) & HARPS (pre/15/20) & HIRES & AAT & RV total & IAD/GOST \\
\midrule
HIP\,8923  & 10/6/6   & \naa/\naa/\naa & \naa & \naa & 22  & 77/49 \\
HIP\,10090 & 12/3/6   & 32/4/\naa    & \naa & \naa & 57  & 106/52 \\
HIP\,21850 & 5/7/11   & 40/4/10     & \naa & 43  & 120 & 108/40 \\
HIP\,39330 & 10/7/10  & 22/3/\naa    & \naa & \naa & 52  & 132/44 \\
HIP\,98599 & 11/\naa/7 & 10/3/\naa    & 5   & \naa & 36  & 143/32 \\
\bottomrule
\end{tabular}
\end{table}


\twocolumn
\section{Tables}\label{sec:appendix_tables}

\begin{table}[ht!]
\caption{\label{tab:hip8923_params}HIP\,8923 joint RV+astrometry parameter estimates.}
\centering
\begin{tabular}{lll}
\toprule
   Parameter            & Prior         & \onek \\
\midrule
\midrule
\multicolumn{3}{c}{Offsets and Jitter} \\
\midrule

$\gamma_{\mathrm{C07}}$~(\ms)  &\Uniform{-200}{200}  &$-30.22^{+6.84}_{-2.18}$ \\
$\gamma_{\mathrm{C14}}$~(\ms)  &\Uniform{-200}{200}  &$-38.17^{+6.77}_{-7.18}$ \\
$\gamma_{\mathrm{C14_T}}$~(\ms)&\Uniform{-200}{200}  & $29.09^{+5.18}_{-2.66}$ \\

$\sigma_{\mathrm{C07}}$~(\ms) &\Normal{0}{10}       &  $1.71^{+1.89}_{-1.71}$ \\
$\sigma_{\mathrm{C14}}$~(\ms) &\Normal{0}{5}        &  $2.82^{+0.79}_{-0.27}$ \\
$\sigma_{\mathrm{C14_T}}$~(\ms)&\Normal{0}{10}       &  $0.94^{+0.83}_{-0.94}$ \\

\midrule
\multicolumn{3}{c}{Keplerian Orbit} \\
\midrule

$P_1$~(days)     & \Uniform{1}{15000}&$5160^{+150}_{-240}$ \\
$K_1$~(\ms)      & \Uniform{0}{200}  &      $48.3^{+3.6}_{-5.5}$ \\
$\phi_1$(rad)    & \Uniform{0}{2\pi} &    $4.96^{+0.65}_{-0.72}$ \\
$e_1$            & \Normal{0}{0.3}    & $0.011^{+0.032}_{-0.011}$ \\
$\omega_1$~(rad)& \Uniform{0}{2\pi}&    $0.92^{+1.14}_{-0.56}$ \\
$I_1$~($^\circ$)    & \Isotro{0}{180}   &   $24.8^{+1.2}_{-3.9}$ \\
$\Omega_1$~($^\circ$)& \Uniform{0}{2\pi} &   $98.47^{+5.40}_{-8.12}$ \\

\midrule
\multicolumn{3}{c}{Astrometry} \\
\midrule

$\Delta \alpha_*$~(mas) &\Uniform{-10}{10} & $0.604^{+0.094}_{-0.009}$ \\
$\Delta\delta$~(mas) &\Uniform{-10}{10} & $0.408^{+0.079}_{-0.068}$ \\
$\Delta \varpi$~(mas) &\Uniform{-10}{10} & $0.017^{+0.017}_{-0.002}$ \\
$\Delta \mu_{\alpha_*}$~(mas/yr) &\Uniform{-10}{10} & $0.220^{+0.021}_{-0.010}$ \\
$\Delta \mu_{\delta}$~(mas/yr) &\Uniform{-10}{10} & $-0.237^{+0.013}_{-0.012}$\\

$J_{\mathrm{hipp}}$~(mas)&\Uniform{0}{10}     &$1.14^{+0.63}_{-0.60}$ \\
$S_{\mathrm{gaia}}$     &\Normal{1}{1, 0.1}  &$1.02^{+0.02}_{-0.06}$ \\

\bottomrule
\end{tabular}
\end{table}

\begin{table}[ht!]
\caption{\label{tab:HIP10090_params}HIP\,10090 joint RV+astrometry parameter estimates.}
\centering
\begin{tabular}{llllll}
\toprule
   Parameter            & Prior         & \onek~             & Parameter            & Prior         & \onek~             \\
\midrule
\midrule
\multicolumn{3}{c}{Offsets and Jitter} & \multicolumn{3}{c}{Astrometry}\\
\midrule

$\gamma_{\mathrm{C07}}$~(\ms)  &\Uniform{-200}{200} &$-24.98^{+6.37}_{-3.68}$ & $\Delta \alpha_*$~(mas)          &\Uniform{-10}{10} & $-0.158^{+0.049}_{-0.036}$ \\
$\gamma_{\mathrm{C14T}}$~(\ms) &\Uniform{-200}{200} & $-4.08^{+7.19}_{-5.16}$ & $\Delta\delta$~(mas)             &\Uniform{-10}{10} & $0.013^{+0.030}_{-0.023}$ \\
$\gamma_{\mathrm{C14}}$~(\ms)  &\Uniform{-200}{200} &  $7.43^{+7.80}_{-7.57}$ & $\Delta \varpi$~(mas)            &\Uniform{-10}{10} &$-0.003^{+0.013}_{-0.003}$ \\
$\gamma_{\mathrm{HIR}}$~(\ms)  &\Uniform{-200}{200} &  $8.52^{+1.52}_{-1.64}$ & $\Delta \mu_{\alpha_*}$~(mas/yr) &\Uniform{-10}{10} &$-0.059^{+0.016}_{-0.001}$ \\
$\gamma_{\mathrm{H15}}$~(\ms) &\Uniform{-200}{200}  &$-17.47^{+1.83}_{-0.65}$ & $\Delta \mu_{\delta}$~(mas/yr)   &\Uniform{-10}{10} &$-0.125^{+0.012}_{-0.010}$ \\

$\sigma_{\mathrm{C07}}$~(\ms)  &\Normal{0}{10}& $2.52^{+2.08}_{-2.52}$ & $J_{\mathrm{hipp}}$~(mas)        &\Uniform{0}{10}   &$1.22^{+0.70}_{-0.13}$ \\
$\sigma_{\mathrm{C14T}}$~(\ms) &\Normal{0}{10}&$1.51^{+1.29}_{-1.15}$  & $S_{\mathrm{gaia}}$              &\Normal{1}{0.1}   &$1.24^{+0.07}_{-0.07}$ \\
$\sigma_{\mathrm{C14}}$~(\ms)  &\Normal{0}{5} &$4.40^{+2.91}_{-4.40}$  & & & \\
$\sigma_{\mathrm{HIR}}$~(\ms)  &\Normal{0}{5} & $4.74^{+0.33}_{-0.23}$ & & & \\
$\sigma_{\mathrm{H15}}$~(\ms) &\Normal{0}{5}  &$0.50^{+0.64}_{-0.50}$  & & & \\

\midrule
\multicolumn{6}{c}{Keplerian Orbits} \\
\midrule

$P_1$~(days)     & \Uniform{1}{10000}   &   $2960^{+120}_{-100}$   & $P_2$~(days)     & \Uniform{1}{10000}& $321.82^{+0.32}_{-0.56}$ \\
$K_1$~(\ms)      & \Uniform{0}{100}     &$23.89^{+1.86}_{-2.86}$   & $K_2$~(\ms)      & \Uniform{0}{100}  &  $11.76^{+1.00}_{-0.22}$ \\
$\phi_1$~(rad)   & \Uniform{0}{2\pi}    & $2.88^{+0.19}_{-0.24}$   & $\phi_2$~(rad)   & \Uniform{0}{2\pi} &   $4.45^{+0.44}_{-0.21}$ \\
$e_1$            & \Normal{0}{0.3}      &$0.115^{+0.005}_{-0.033}$ & $e_2$            & \Normal{0}{0.3}   &$0.188^{+0.043}_{-0.048}$ \\
$\omega_1$~(rad) & \Uniform{0}{2\pi}    &    $1.82^{+0.32}_{-0.15}$& $\omega_2$~(rad) & \Uniform{0}{2\pi} &   $4.94^{+0.24}_{-0.33}$ \\
$I_1$~($^\circ$) & \Isotro{0}{180}      &   $156.5^{+1.8}_{-2.2}$  & $I_2$~($^\circ$) & \Isotro{0}{180}   & $153.8^{+1.3}_{-3.1}$     \\
$\Omega_1$~($^\circ$)& \Uniform{0}{2\pi}&    $341.63^{+4.04}_{-0.01}$ & $\Omega_2$~($^\circ$) & \Uniform{0}{2\pi} & $341.87^{+2.99}_{-1.28}$ \\

\bottomrule
\end{tabular}
\end{table}

\begin{table}
\caption{\label{tab:HIP98599_params}HIP\,98599 joint RV+astrometry parameter estimates.}
\centering
\begin{tabular}{lll}
\toprule
   Parameter            & Prior         & \onek~ \\
\midrule
\midrule
\multicolumn{3}{c}{Offsets and Jitter} \\
\midrule

$\gamma_{\mathrm{C07}}$~(\ms)  &\Uniform{-200}{200} &$-42.12^{+2.61}_{-3.63}$ \\
$\gamma_{\mathrm{C14}}$~(\ms)  &\Uniform{-200}{200} & $15.58^{+2.82}_{-4.41}$ \\
$\gamma_{\mathrm{H}}$~(\ms) &\Uniform{-200}{200} & $-1.09^{+1.58}_{-1.53}$ \\
$\gamma_{\mathrm{HIR}}$~(\ms)  &\Uniform{-200}{200} & $28.84^{+8.58}_{-9.66}$ \\
$\gamma_{\mathrm{H15}}$~(\ms) &\Uniform{-200}{200} & $26.08^{+0.72}_{-2.45}$ \\

$\sigma_{\mathrm{C07}}$~(\ms)  &\Normal{0}{10} &$3.63^{+3.41}_{-3.63}$ \\
$\sigma_{\mathrm{C14}}$~(\ms)  &\Normal{0}{5}  &$0.21^{+0.55}_{-0.21}$ \\
$\sigma_{\mathrm{H}}$~(\ms) &\Normal{0}{5}  &$6.90^{+1.53}_{-0.27}$ \\
$\sigma_{\mathrm{HIR}}$~(\ms)  &\Normal{0}{10} &$9.91^{+1.50}_{-1.59}$ \\
$\sigma_{\mathrm{H15}}$~(\ms) &\Normal{0}{5}  &$2.56^{+1.52}_{-1.30}$ \\

\midrule
\multicolumn{3}{c}{Keplerian Orbit} \\
\midrule

$P_1$~(days)     & \Uniform{1}{10000}& $2656^{+40}_{-16}$ \\
$K_1$~(\ms)      & \Uniform{0}{150}  &  $45.53^{+1.04}_{-2.90}$ \\
$\phi_1$~(rad)   & \Uniform{0}{2\pi} &   $5.47^{+0.25}_{-0.05}$ \\
$e_1$            & \Normal{0}{0.3}    &$0.169^{+0.002}_{-0.055}$ \\
$\omega_1$~(rad)& \Uniform{0}{2\pi} &   $4.80^{+0.06}_{-0.28}$ \\
$I_1$~($^\circ$)     & \Isotro{0}{180}   &  $29.7^{+6.9}_{-6.2}$ \\
$\Omega_1$~($^\circ$)& \Uniform{0}{2\pi} & $302.26^{+7.55}_{-1.16}$ \\

\midrule
\multicolumn{3}{c}{Astrometry} \\
\midrule

$\Delta \alpha_*$~(mas)       &\Uniform{-10}{10} &$-0.164^{+0.022}_{-0.020}$ \\
$\Delta\delta$~(mas)          &\Uniform{-10}{10} & $0.151^{+0.016}_{-0.043}$ \\
$\Delta \varpi$~(mas)         &\Uniform{-10}{10} & $-0.028^{+0.028}_{-0.023}$\\
$\Delta \mu_{\alpha_*}$~(mas/yr) &\Uniform{-10}{10} & $0.106^{+0.015}_{-0.008}$ \\
$\Delta \mu_{\delta}$~(mas/yr)   &\Uniform{-10}{10} & $0.138^{+0.016}_{-0.001}$ \\

$J_{\mathrm{hipp}}$~(mas)&\Uniform{0}{10}   &$0.67^{+0.57}_{-0.50}$ \\
$S_{\mathrm{gaia}}$      &\Normal{1}{1, 0.1}&$1.24^{+0.03}_{-0.03}$ \\

\bottomrule
\end{tabular}
\end{table}

\onecolumn
\section{More Tables}\label{sec:appendix_more_tables}

\begin{table}[ht!]
\caption{\label{tab:HIP39330_params}HIP\,39330 joint RV+astrometry parameter estimates. Models \twok$_1$ and \twok$_2$ have different inclination priors.}
\centering
\begin{tabular}{lllll}
\toprule
   Parameter            & Prior         & \twok & \twok$_1$& \twok$_2$\\
\midrule
\midrule
\multicolumn{5}{c}{Offsets and Jitter} \\
\midrule

$\gamma_{\mathrm{C07}}$~(\ms)  &\Uniform{-200}{200} &$14.29^{+0.44}_{-2.88}$ &$13.37^{+3.55}_{-3.70}$ &$16.64^{+3.51}_{-5.88}$  \\
$\gamma_{\mathrm{C14}}$~(\ms)  &\Uniform{-200}{200} &$17.60^{+0.45}_{-4.10}$ &$14.76^{+2.08}_{-2.28}$ &$11.87^{+2.77}_{-1.46}$  \\
$\gamma_{\mathrm{C14T}}$~(\ms) &\Uniform{-200}{200} &$-5.42^{+1.55}_{-3.72}$  &$-6.95^{+1.30}_{-3.82}$ &$-0.57^{+5.22}_{-5.30}$  \\
$\gamma_{\mathrm{H}}$~(\ms)   &\Uniform{-200}{200} &$8.95^{+2.92}_{-0.20}$  &$12.59^{+3.15}_{-3.64}$ &  $8.35^{+0.42}_{-2.27}$ \\
$\gamma_{\mathrm{H15}}$~(\ms)   &\Uniform{-200}{200} &$-13.16^{+4.94}_{-5.09}$&$-16.84^{+1.99}_{-2.29}$&$-13.76^{+3.14}_{-0.98}$ \\

$\sigma_{\mathrm{C07}}$~(\ms)  &\Normal{0}{10}&$2.30^{+2.35}_{-2.30}$&  $1.13^{+0.68}_{-1.13}$& $1.82^{+0.08}_{-1.82}$ \\
$\sigma_{\mathrm{C14}}$~(\ms)  &\Normal{0}{5} &$3.69^{+0.71}_{-0.31}$&  $3.86^{+0.85}_{-0.28}$& $3.45^{+0.95}_{-0.09}$ \\
$\sigma_{\mathrm{C14T}}$~(\ms) &\Normal{0}{10}&$4.78^{+3.03}_{-4.77}$&  $3.98^{+2.99}_{-3.98}$& $2.22^{+0.18}_{-2.21}$ \\
$\sigma_{\mathrm{H}}$~(\ms)   &\Normal{0}{5} &$5.09^{+0.42}_{-0.30}$&  $4.69^{+0.90}_{-0.79}$& $4.93^{+0.65}_{-0.07}$ \\
$\sigma_{\mathrm{H15}}$~(\ms)   &\Normal{0}{5} &$3.56^{+1.13}_{-0.65}$&  $4.45^{+0.48}_{-1.40}$& $2.82^{+2.32}_{-1.93}$ \\

\midrule
\multicolumn{5}{c}{Keplerian Orbit} \\
\midrule

$P_1$~(days)          & \Uniform{1}{10000}&$4596.9^{+1.7}_{-154.0}$ &$4654.2^{+211.5}_{-248.4}$&$4490.5^{+29.8}_{-144.1}$ \\
$K_1$~(\ms)           & \Uniform{0}{100}  &$20.96^{+1.28}_{-2.04}$  &   $19.78^{+1.25}_{-1.80}$&  $18.28^{+0.12}_{-2.53}$ \\
$\phi_1$~(rad)        & \Uniform{0}{2\pi} &$6.12^{+0.16}_{-0.51}$   &    $5.05^{+1.15}_{-1.33}$&   $5.88^{+0.40}_{-0.08}$ \\
$e_1$                 & \Normal{0}{0.3}    &$0.109^{+0.053}_{-0.109}$& $0.078^{+0.053}_{-0.078}$&$0.064^{+0.063}_{-0.064}$ \\
$\omega_1$~(rad)& \Uniform{0}{2\pi} &$1.52^{+0.71}_{-0.17}$   &    $2.85^{+0.05}_{-1.21}$&   $1.69^{+0.23}_{-0.41}$ \\
$I_1$~($^\circ$)      & \Isotro{0}{180}   &$110.10^{+4.95}_{-5.72}$ &   $55.01^{+0.55}_{-3.86}$& $127.14^{+3.38}_{-1.24}$ \\
$\Omega_1$~($^\circ$) & \Uniform{0}{2\pi} &$5.62^{+4.91}_{-5.22}$   &  $253.32^{+3.56}_{-1.12}$&   $4.05^{+4.96}_{-5.05}$ \\
\midrule

$P_2$~(days)          & \Uniform{1}{10000}&$257.11^{+1.55}_{-1.41}$ & $258.34^{+0.90}_{-0.37}$& $258.75^{+0.29}_{-0.99}$ \\
$K_2$~(\ms)           & \Uniform{0}{100}  &$8.10^{+1.56}_{-1.31}$   &   $9.73^{+0.25}_{-1.03}$&  $10.44^{+1.41}_{-1.72}$ \\
$\phi_2$~(rad)        & \Uniform{0}{2\pi} &$3.45^{+0.04}_{-1.75}$   &   $2.33^{+1.24}_{-0.35}$&   $3.08^{+0.27}_{-1.31}$ \\
$e_2$                 & \Normal{0}{0.3}    &$0.020^{+0.002}_{-0.020}$&$0.079^{+0.053}_{-0.079}$&$0.099^{+0.053}_{-0.099}$ \\
$\omega_2$~(rad)& \Uniform{0}{2\pi} &$2.69^{+2.05}_{-1.66}$   &   $4.01^{+0.52}_{-1.08}$&   $3.32^{+1.46}_{-0.14}$ \\
$I_2$~($^\circ$)      & \Isotro{0}{180}   &$110.02^{+5.13}_{-5.31}$ &  $53.77^{+1.54}_{-3.33}$& $132.26^{+1.44}_{-3.46}$ \\
$\Omega_2$~($^\circ$) & \Uniform{0}{2\pi} &$1.67^{+3.89}_{-1.00}$   & $256.48^{+3.68}_{-1.21}$   &$5.96^{+5.15}_{-8.33}$ \\

\midrule
\multicolumn{5}{c}{Astrometry} \\
\midrule

$\Delta \alpha_*$~(mas)       &\Uniform{-10}{10} & $0.083^{+0.029}_{-0.045}$&$-0.24^{+0.056}_{-0.053}$& $0.143^{+0.039}_{-0.039}$ \\
$\Delta\delta$~(mas)          &\Uniform{-10}{10} & $0.159^{+0.052}_{-0.050}$&$0.006^{+0.022}_{-0.013}$& $0.178^{+0.020}_{-0.027}$ \\
$\Delta \varpi$~(mas)         &\Uniform{-10}{10} &$-0.005^{+0.015}_{-0.001}$&$0.006^{+0.004}_{-0.013}$& $0.009^{+0.003}_{-0.013}$ \\
$\Delta \mu_{\alpha_*}$~(mas/yr)&\Uniform{-10}{10}& $0.005^{+0.012}_{-0.016}$&$-0.034^{+0.013}_{-0.012}$&$-0.008^{+0.002}_{-0.010}$ \\
$\Delta \mu_{\delta}$~(mas/yr)  &\Uniform{-10}{10}& $0.057^{+0.002}_{-0.009}$&$0.054^{+0.003}_{-0.010}$& $0.067^{+0.012}_{-0.017}$ \\

$J_{\mathrm{hipp}}$~(mas)&\Uniform{0}{10}   &$0.16^{+0.12}_{-0.16}$&$0.60^{+0.33}_{-0.60}$&  $0.70^{+0.32}_{-0.70}$ \\
$S_{\mathrm{gaia}}$      &\Normal{1}{0.1}   &$1.24^{+0.11}_{-0.07}$&$1.32^{+0.03}_{-0.04}$&  $1.30^{+0.06}_{-0.01}$ \\

\bottomrule
\end{tabular}
\end{table}

\begin{table*}[ht!]
\centering
\caption{\label{tab:planets-summary1}Planet signatures summary, with Jupiter analogues in bold.}
\begin{tabular}{lllllll}
\toprule
\multicolumn{1}{c}{} & \multicolumn{2}{c}{Period (days)} & \multicolumn{2}{c}{Mass ($M_{J}$)} & \multicolumn{2}{c}{Semi-major axis (AU)}\\
\midrule
Planet    &RV  &RV+AM &RV  &RV+AM &RV  &RV+AM\\
\midrule
\midrule
HIP\,8923b   &$5116_{-259}^{+456}$&$5160^{+150}_{-240}$ &$4.30_{-0.75}^{+0.07}$&$9.98_{-0.16}^{+0.78}$ & $5.89_{-0.26}^{+0.30}$&$5.90_{-0.22}^{+0.10}$ \\
HIP\,10090b  &$2975_{-97}^{+139}$&$2960^{+120}_{-100}$&$3.95_{-0.10}^{+0.14}$&$3.87_{-0.60}^{+0.65}$ & $1.53_{-0.14}^{+0.12}$&$3.87_{-0.11}^{+0.19}$ \\
HIP\,10090c  &$321.3_{-0.4}^{+0.5}$&$321.8^{+0.3}_{-0.6}  $&$0.39_{-0.03}^{+0.02}$&$0.85_{-0.12}^{+0.03}$ &$0.87_{-0.01}^{+0.03}$&$0.90_{-0.01}^{+0.01}$\\

HIP\,21850b &$2540^{+4}_{-5}$     &$2539^{+1}_{-3}       $ & $8.21^{+0.13}_{-0.46}$&$8.25_{-0.85}^{+1.05}$ &$3.65_{-0.11}^{+0.02}]$ &$3.49_{-0.14}^{+0.18}$\\
HIP\,21850c &$15903^{+2470}_{-4840}$&$11320^{+490}_{-940}$ & $6.62^{+1.37}_{-2.46}$&$4.67_{-0.43}^{+0.44}$ & $12.45_{-2.71}^{+1.40}$&$9.90_{-0.62}^{+0.29}$\\

HIP\,39330b  &$4636_{-197}^{+300}$&$4650^{+210}_{-250}  $& $2.59_{-0.07}^{+0.51}$&$1.67_{-0.08}^{+0.19}$ &$5.43_{-0.17}^{+0.16}$&$5.05_{-0.17}^{+0.23}$ \\

HIP\,98599b  & $2579_{-90}^{+79}$&$2656^{+40}_{-16}$ & $3.28_{-0.13}^{+ 0.17}$&$6.85_{-0.22}^{+0.10}$ & $3.95_{-0.05}^{+ 0.03}$&$3.96_{-0.01}^{+0.07}$ \\
\bottomrule
\end{tabular}
\end{table*}






\end{appendix}
\end{document}